\documentclass[prb,tighten]{revtex4}

\usepackage{epsfig}

\usepackage{graphicx}

\begin{document}

\title{Mass and heat fluxes for a binary granular mixture at low-density}
\author{Vicente Garz\'{o}}
\email[E-mail: ]{vicenteg@unex.es}
\address{Departamento de F\'{\i}sica, Universidad de Extremadura, E-06071
Badajoz, Spain}
\author{Jos\'e Mar\'{\i}a Montanero}
\email[E-mail: ]{jmm@unex.es}
\address{Departamento de Electr\'onica e Ingenier\'{\i}a Electromec\'anica,
Universidad de Extremadura, E-06071 Badajoz, Spain}
\author{James W. Dufty}
\email[E-mail: ]{dufty@phys.ufl.edu}
\address{Department of Physics, University of Florida, Gainesville, Florida
32611}

\begin{abstract}

The Navier--Stokes order hydrodynamic equations for a low density
granular mixture obtained previously from the Chapman--Enskog
solution to the Boltzmann equation are considered further. The six
transport coefficients associated with mass and heat flux in a
binary mixture are given as functions of the mass ratio, size
ratio, composition, and coefficients of restitution. Their
quantitative variation across this parameter set is demonstrated
using low order Sonine polynomial approximations to solve the
exact integral equations. The results are also used to quantify
the violation of the Onsager reciprocal relations for a granular
mixture. Finally, the stability of the homogeneous cooling state
is discussed.

\end{abstract}

\date{\today}
\maketitle

\draft

\section{Introduction}
\label{sec1}

The relevance and context of a hydrodynamic description for
granular gases remains a controversial topic. At sufficiently low
density the origin of hydrodynamics can be studied from Boltzmann
kinetic theory. \cite{GS95,BP04} To obtain hydrodynamics from the
Boltzmann kinetic equation the essential assumption is the
existence of a ``normal'' solution, defined to be one for which
all space and time dependence occurs through the macroscopic
hydrodynamic fields. \cite{CC70} This solution, together with the
macroscopic balance equations, leads to a closed set of
hydrodynamic equations for these fields. The Chapman--Enskog
method provides a constructive means to obtain an approximation to
such a solution for states whose spatial gradients are not too
large. In this general context, the study of hydrodynamics for
granular gases is the same as that for normal gases.

The details of the Chapman--Enskog method have been carried out
for a one granular component gas to obtain the Navier--Stokes
order hydrodynamic equations, together with exact integral
equations determining the transport coefficients occurring in
these equations. \cite{BDKS98} These integral equations have been
solved approximately using Sonine polynomial expansions and
compared to both Direct Monte Carlo simulation of the Boltzmann
equation and molecular dynamics simulation of the gas.
\cite{BRMC99} Good agreement is obtained even for relatively
strong degrees of dissipation. The results support the formal
theoretical analysis and the claim that hydrodynamics is not
limited to the quasi-elastic limit.

The analysis for multicomponent granular gases is much more
complicated than for a one component gas. Most of the previous
attempts \cite{JM89} were made for {\em nearly} elastic spheres
where the equipartition of energy can be considered as an
acceptable assumption. In addition, according to this level of
approximation, the inelasticity is only accounted for by the
presence of a sink term in the energy balance equation, so that
the expressions for the transport coefficients are the same as
those obtained for normal fluids. \cite{MCK83} However, the
failure of energy equipartition in multicomponent granular gases
\cite{GD99} has been also confirmed by computer simulations
\cite{computer} and even observed in real experiments. \cite{exp}
Although the possibility of nonequipartition was already pointed
out many years ago, \cite{JM87} it has not been until recently
that a systematic study of the effect of nonequipartition on
transport has been carefuly analyzed. In this context, Garz\'o and
Dufty \cite{GD02} have carried out a derivation of the
Navier--Stokes hydrodynamic equations for a binary mixture at
low-density that accounts for nonequipartition of granular energy.
These equations and associated transport coefficients provide a
somewhat more stringent test of the analysis since the parameter
space is much larger. There are now many more transport
coefficients, given as functions of the three independent
coefficients of restitution, size ratio, mass ratio, and
composition. As in the one component case, explicit expressions
for the transport coefficients requires also to consider Sonine
polynomial expansions. The accuracy of this approach has been
confirmed by comparison with Monte Carlo simulations of the
Boltzmann equation in the cases of the shear viscosity \cite{MG03}
and the tracer diffusion \cite{GM04} coefficients. Exceptions to
this agreement are extreme mass or size ratios and strong
dissipation, although these discrepancies between theory and
simulation diminish as one considers more terms in the Sonine
polynomial approximation. \cite{GM04}

Since the dependence of the shear viscosity coefficient on the
parameters of the mixture (masses, sizes, concentration,
coefficients of restitution) has been widely studied in a previous
work, \cite{MG03}  a primary objective here is to demonstrate the
variation of the six transport coefficients associated with the
mass and heat flux in this parameter space, using the same Sonine
polynomial approximation as was found applicable for the one
component gas. To set the context for these quantitative results
two qualitative and potentially confusing issues are briefly noted
at the outset.

There is some ambiguity regarding the hydrodynamic temperature in
a mixture since temperatures for each species can be defined in
addition to the global temperature. What should be the
hydrodynamic fields? For normal gases the answer is clear. These
are set by the slow variables (at large space and time scales)
associated with conserved quantities. In addition to species
number and momentum, only the total kinetic energy is conserved,
so only one global temperature occurs as a hydrodynamic field. The
energy is no longer conserved for granular gases, but it remains a
slow variable if the cooling rate is not too large. Thus, in this
case as well, only the global temperature should appear among the
hydrodynamic fields.

Nevertheless, the species temperatures play a new and interesting
secondary role. For a normal gas, there is a rapid velocity
relaxation in each fluid cell to a local equilibrium state on the
time scale of a few collisions. Subsequently, the equilibration
among cells occurs via the hydrodynamic equations. In each cell
the species velocity distributions are characterized by the
species temperatures. These are approximately the same due to
equipartition, and the hydrodynamic relaxation occurs for the
single common temperature. \cite{CC70} A similar rapid velocity
relaxation occurs for granular gases in each small cell, but to a
universal state different from local equilibrium and one for which
equipartition no longer occurs. Hence, the species temperatures
$T_{i}$ are different from each other and from the overall
temperature $T$ of the cell. \cite{GD99} Nevertheless, the time
dependence of all temperatures is the same in this and subsequent
states, $T_{i}(t)=\gamma _{i}T(t)$. This implies that the species
temperatures do not provide any new dynamical degree of freedom.
They still characterize the shape of the partial velocity
distributions and affect the quantitative averages calculated with
these distributions. The transport coefficients for granular
mixtures therefore have new quantitative effects arising from the
time independent temperature ratio $\gamma _{i}$ for each species.
\cite{GD02} This dependence is illustrated here as well.

In some earlier works, \cite{JM87,GDH05} additional equations for
each species temperature have been included among the hydrodynamic
set. This is possible since the overall temperature is determined
from these by $T(t)=\sum_{i}x_{i}T_{i}(t)$, where $ x_{i}$ is the
concentration of species $i$. However, this is an unnecessary
complication, describing additional kinetics beyond hydrodynamics
that is relevant only on the time scale of a few collisions. As
described above (and supported by molecular dynamics simulations
\cite{DHGD02}) the dynamics of all temperatures quickly reduces to
that of $T(t)$. The remaining time independent determination of
the $\gamma _{i}$ follows directly from the condition that the
cooling rates of all temperatures be the same.

A second confusing issue is the context of the Navier--Stokes
equations considered here. The derivation of the Navier--Stokes
order transport coefficients does not limit their application to
weak inelasticity. For this reason the results reported below
include a domain of both weak and strong inelasticity, $0.5\leq
\alpha \leq 1$. The Navier--Stokes hydrodynamic equations
themselves may or may not be limited with respect to inelasticity,
depending on the particular states considered. The derivation of
these equations by the Chapman--Enskog method assumes that
relative changes in the hydrodynamic fields over distances of the
order of a mean free path are small. For normal gases this can be
controlled by the initial or boundary conditions. It is more
complicated for granular gases. In some cases (e.g., steady states
such as the simple shear flow problem \cite{SGD04}) the boundary
conditions imply a relationship between the coefficient of
restitution and some hydrodynamic gradient -- the two cannot be
chosen independently. Consequently, there are examples for which
the Navier--Stokes approximation is never valid or is restricted
to the quasi-elastic limit. \cite{SGD04} However, the transport
coefficients characterizing the Navier--Stokes order hydrodynamic
equations are well-defined functions of $\alpha$, regardless of
the applicability of those equations (e.g., the coefficients of a
Taylor series for some function $f(\alpha,x)$ in powers of $x$ may
be defined for all $\alpha$ at each order, even when the series
cannot be truncated at that order).

The plan of the paper is as follows. First, in Sec.\ \ref{sec2}
the hydrodynamic equations and associated fluxes to Navier--Stokes
order are recalled, and the expressions for the transport
coefficients for heat and mass transport are given to leading
Sonine polynomial approximation. The elastic limit is discussed to
aid in the interpretation of these expressions. Next, in Sec.\
\ref{sec3} the results for these six coefficients are illustrated
for a common coefficient of restitution  and same size ratio as
functions of $\alpha $ at a composition $x_{1}=0.2$ for several
values of the mass ratio. With the exception of thermal
conductivity, the deviations from normal gas values are largest at
small $\alpha $ and large mass ratio. The usual Onsager relations
among these coefficients for normal gases is then noted and tested
for the granular gas in Sec.\ \ref{sec4}. Since the underlying
basis for these relations (time reversal symmetry) no longer holds
for granular systems, the expected violation is demonstrated as a
function of $\alpha $ for the same conditions. The stability of a
special homogeneous solution to the mixture hydrodynamic equations
is then studied, and some comments on the implications of the
instability found are offered. Finally, the results are summarized
and discussed in the last section.

\section{Boltzmann kinetic theory for the mass and heat fluxes}
\label{sec2}

We consider a binary mixture of \emph{inelastic}, smooth, hard
spheres of masses $m_{1}$ and $m_{2}$, and diameters $\sigma _{1}$
and $\sigma _{2}$. The inelasticity of collisions among all pairs
is characterized by three independent constant coefficients of
normal restitution $\alpha _{11}$, $ \alpha _{22}$, and $\alpha
_{12}=\alpha _{21}$, where $\alpha _{ij}$ is the coefficient of
restitution for collisions between particles of species $i$ and
$j$. \ It is assumed that the density of each species is
sufficiently low that their velocity distribution functions are
accurately described by the coupled set of {\em inelastic}
Boltzmann kinetic equations. The precise form of these equations
is given in Ref.\ \onlinecite{GD02} and will not be required here.
These equations imply the exact macroscopic balance equations for
the particle number density of each species, $n_{i}\left(
\mathbf{r},t\right) $, flow velocity $\mathbf{u}\left(
\mathbf{r},t\right) $, and temperature, $T\left(
\mathbf{r},t\right) $, \cite{GD02}
\begin{equation}
D_{t}n_{i}+n_{i}\nabla \cdot \mathbf{u}+\frac{\nabla \cdot
\mathbf{j}_{i}}{ m_{i}}=0\;,\hspace{0.3in}i=1,2  \label{2.01}
\end{equation}
\begin{equation}
D_{t}\mathbf{u}+\rho ^{-1}\nabla \mathsf{P}=0\;,  \label{2.02}
\end{equation}
\begin{equation}
D_{t}T-\frac{T}{n}\sum_{i}\frac{\nabla \cdot
\mathbf{j}_{i}}{m_{i}}+\frac{2}{ 3n}\left( \nabla \cdot
\mathbf{q}+\mathsf{P}:\nabla \mathbf{u}\right) =-\zeta T\;.
\label{2.03}
\end{equation}
In the above equations, $D_{t}=\partial _{t}+\mathbf{u}\cdot
\nabla $ is the material derivative, $\rho =m_{1}n_{1}+m_{2}n_{2}$
is the total mass density, $\mathbf{j}_{i}$ is the particle number
flux for species $i$, $\mathbf{q}$ is the heat flux, $\mathsf{P}$
is the pressure tensor, and $\zeta$ is the cooling rate. For the
two component mixture considered here there are six independent
fields, $n_{1},$ $n_{2},$ $T,$ $\mathbf{u}$. To obtain a closed
set of hydrodynamic equations, expressions for ${\bf j}_{i}$,
$\mathbf{q}$, $ \mathsf{P}$, and $\zeta $ must be given in terms
of these fields. Such expressions are called ``constitutive
equations''. It is convenient to give these constitutive equations
in terms of a different set of experimentally more accessible
fields, $x_{1},$ $p,$ $T,$ $\mathbf{u}$, where $
x_{1}=n_{1}/\left( n_{1}+n_{2}\right) $ is the composition of
species $1$, and $p=\left( n_{1}+n_{2}\right) T$ is the
hydrostatic pressure. This is simply a change of variables, so
that Eqs.\ (\ref{2.01})--(\ref{2.03}) become
\begin{equation}
D_{t}x_{1}+\frac{\rho }{n^{2}m_{1}m_{2}}\nabla \cdot
\mathbf{j}_{1}=0\;, \label{2.04}
\end{equation}
\begin{equation}
D_{t}p+p\nabla \cdot \mathbf{u}+\frac{2}{3}\left( \nabla \cdot
\mathbf{q}+ \mathsf{P}:\nabla \mathbf{u}\right) =-\zeta p,
\label{2.05}
\end{equation}
\begin{equation}
D_{t}\mathbf{u}+\rho ^{-1}\nabla \mathsf{P}=0\;,  \label{2.06}
\end{equation}
\begin{equation}
D_{t}T-\frac{T}{n}\sum_{i}\frac{\nabla \cdot
\mathbf{j}_{i}}{m_{i}}+\frac{2}{3n}\left( \nabla \cdot
\mathbf{q}+\mathsf{P}:\nabla \mathbf{u}\right) =-\zeta T\;.
\label{2.07}
\end{equation}

The constitutive equations up to the Navier--Stokes order have
been obtained from the Boltzmann equation in Ref.\
\onlinecite{GD02} with the results
\begin{equation}
\mathbf{j}_{1}=-\left( \frac{m_{1}m_{2}n}{\rho }\right) D\nabla
x_{1}-\frac{ \rho }{p}D_{p}\nabla p-\frac{\rho }{T}D^{\prime
}\nabla T,\quad \mathbf{j}_{2}=-\mathbf{j}_{1},  \label{2.1}
\end{equation}
\begin{equation}
\mathbf{q}=-T^{2}D^{\prime \prime }\nabla x_{1}-L\nabla p-\lambda
\nabla T, \label{2.2}
\end{equation}
\begin{equation}
P_{k\ell }=p\delta _{k\ell }-\eta \left( \nabla _{\ell
}u_{k}+\nabla _{k}u_{\ell }-\frac{2}{3}\delta _{k\ell }\nabla
\cdot \mathbf{u}\right) , \label{2.3}
\end{equation}
\begin{equation}
\label{2.2.1} \zeta =\zeta _{0}+\mathcal{O}(\nabla^2)
\end{equation}
The transport coefficients $\{D,D_{p},D^{\prime},D^{\prime \prime
},L,\lambda ,\eta \}$ verify a set of coupled linear integral
equations which can be solved approximately by using the leading
terms in a Sonine polynomial expansion. This solution provides
explicit expressions for the transport coefficients in terms of
the coefficients of restitution and the parameters of the mixture
(masses, sizes, and composition). The above expressions for mass
and heat fluxes can be defined in a variety of equivalent ways
depending on the choice of driving forces used. For systems with
elastic collisions, the specific set of gradients contributing to
each flux is restricted by fluid symmetry, Onsager's relations
(time reversal invariance), and the form of entropy production.
\cite{GM84} In this case, one usual representation leads to the
mass and heat fluxes proportional to $(\nabla \mu_{i})_T$ and
$\nabla T$, where $\mu_i$ (defined below) is the chemical
potential per unit mass. However, for \emph{inelastic} systems
only fluid symmetry holds and so there is more flexibility in
representing the fluxes and identifying the corresponding
transport coefficients. In particular, a third contribution
proportional to $\nabla p$ appears in both fluxes. Some care is
required in comparing transport coefficients in different
representations using different independent gradients for the
driving forces.

The cooling rate to lowest order in the gradients is $\zeta =\zeta
_{0}(x_{1},p,T)$. There are no contributions to first order in the
gradient for the low density Boltzmann equation. The general form
including second order gradient contributions is displayed in
Appendix A. These second order terms have been calculated for a
one component gas \cite{BDKS98} and found to be very small. Here,
these second order contributions to the cooling rate will be
neglected.

Substitution of the Navier--Stokes constitutive equations, (\ref
{2.1})-(\ref{2.3}), into the exact balance equations,
(\ref{2.04})-(\ref{2.07} ), gives the Navier--Stokes hydrodynamic
equations for a binary mixture
\begin{equation}
D_{t}x_{1}=\frac{\rho }{n^{2}m_{1}m_{2}}\nabla \cdot \left(
\frac{m_{1}m_{2}n }{\rho }D\nabla x_{1}+\frac{\rho }{p}D_{p}\nabla
p+\frac{\rho }{T}D^{\prime }\nabla T\right) \;,  \label{2.4}
\end{equation}
\begin{eqnarray}
\left( D_{t}+\zeta \right) p+\frac{5}{3}p\nabla \cdot \mathbf{u}
&=&\frac{2}{ 3}\nabla \cdot \left( T^{2}D^{\prime \prime }\nabla
x_{1}+L\nabla
p+\lambda\nabla T\right)   \nonumber \\
&&+\frac{2}{3}\eta \left( \nabla _{\ell }u_{k}+\nabla _{k}u_{\ell
}-\frac{2}{ 3}\delta _{k\ell }\nabla \cdot \mathbf{u}\right)
\nabla _{\ell }u_{k}, \label{2.5}
\end{eqnarray}
\begin{eqnarray}
\left( D_{t}+\zeta \right) T+\frac{2}{3}p\nabla \cdot \mathbf{u}
&=&-\frac{T }{n}\frac{m_{2}-m_{1}}{m_{1}m_{2}}\nabla \cdot \left(
\frac{m_{1}m_{2}n}{ \rho }D\nabla x_{1}+\frac{\rho }{p}D_{p}\nabla
p+\frac{\rho
}{T}D^{\prime}\nabla T\right)   \nonumber \\
&&+\frac{2}{3n}\nabla \cdot \left( T^{2}D^{\prime \prime }\nabla
x_{1}+L\nabla p+\lambda \nabla T\right)   \nonumber \\
&&+\frac{2}{3n}\eta \left( \nabla _{\ell }u_{k}+\nabla _{k}u_{\ell
}-\frac{2 }{3}\delta _{k\ell }\nabla \cdot \mathbf{u}\right)
\nabla _{\ell }u_{k}, \label{2.6}
\end{eqnarray}
\begin{equation}
D_{t}u_{\ell }+\rho ^{-1}\nabla _{\ell }p=\rho ^{-1}\nabla
_{k}\eta \left( \nabla _{\ell }u_{k}+\nabla _{k}u_{\ell
}-\frac{2}{3}\delta _{k\ell }\nabla \cdot \mathbf{u}\right) \;.
\label{2.7}
\end{equation}
For the chosen set of fields $n=p/T$ and $\rho =p\left[ \left(
m_{1}-m_{2}\right) x_{1}+m_{2}\right] /T$. These equations are
exact to second order in the spatial gradients for a low density
Boltzmann gas.

\subsection{Mass flux}
\label{subsec2.1}

The mass flux contains three transport coefficients, the diffusion
coefficient $D$, the pressure diffusion coefficient $D_{p}$, and
the thermal diffusion coefficient $D^{\prime }$. Explicit
expressions for these were obtained in Ref.\ \onlinecite{GD02}
using a first Sonine approximation. Dimensionless forms are
defined by

\begin{equation}
D=\frac{\rho T}{m_{1}m_{2}\nu _{0}}D^{\ast },\quad
D_{p}=\frac{nT}{\rho \nu _{0}}D_{p}^{\ast },\quad D^{\prime
}=\frac{nT}{\rho \nu _{0}}D^{\prime}{}^{\ast }.  \label{2.8}
\end{equation}
Here, $\nu _{0}=\sqrt{\pi }n\sigma _{12}^{2}v_{0}$,
$\sigma_{12}=(\sigma_1+\sigma_2)/2$, and  $v_{0}=\sqrt{
2T(m_{1}+m_{2})/m_{1}m_{2}}$ is a thermal velocity defined in
terms of the temperature $T$ of the mixture. The explicit forms
are then
\begin{equation}
D^{\ast }=\left[ \left( \frac{\partial }{\partial
x_{1}}x_{1}\gamma _{1}\right) _{p,T}+\left( \frac{\partial \zeta
^{\ast }}{\partial x_{1}} \right) _{p,T}\left( 1-\frac{\zeta
^{\ast }}{2\nu ^{\ast }}\right) D_{p}^{\ast }\right] \left( \nu
^{\ast }-\frac{1}{2}\zeta ^{\ast }\right)^{-1},  \label{2.9}
\end{equation}
\begin{equation}
D_{p}^{\ast }=x_{1}\left[ \gamma _{1}-\frac{\mu (1+\delta )}{1+\mu
\delta } \right] \left( \nu ^{\ast }-\frac{3}{2}\zeta ^{\ast
}+\frac{\zeta ^{\ast 2}}{ 2\nu ^{\ast }}\right) ^{-1},
\label{2.10}
\end{equation}
\begin{equation}
D^{\prime }{}^{\ast }=-\frac{\zeta ^{\ast }}{2\nu ^{\ast
}}D_{p}^{\ast }. \label{2.11}
\end{equation}
In these equations,
\begin{equation}
\gamma _{1}=\frac{T_{1}}{T}=\frac{\gamma }{1+x_{1}(\gamma
-1)},\quad \gamma _{2}=\frac{T_{2}}{T}=\frac{1}{1+x_{1}(\gamma
-1)},  \label{2.11a}
\end{equation}
where $\mu = m_{1}/m_{2}$ is the mass ratio, $\mu
_{ij}=m_{i}/(m_{i}+m_{j}),$ $\delta = x_{1}/x_{2}$, and $\gamma
=T_{1}/T_{2}$. The detailed forms for the temperature ratio
$\gamma ,$ dimensionless collision rate $\nu ^{\ast }$, and
dimensionless cooling rate $\zeta ^{\ast }$ are given in Appendix
A. Since $\mathbf{j}_{1}=-\mathbf{j}_{2}$ and $\nabla
x_{1}=-\nabla x_{2}$, it is expected that $D^{\ast }$ should be
symmetric with respect to interchange of particles $1$ and $2$
while $D_{p}^{\ast }$ and $D^{\prime }{}^{\ast}$ should be
antisymmetric. This can be easily verified by noting that $
x_{1}\gamma _{1}+x_{2}\gamma _{2}=1$.

In the case of elastic collisions $\alpha _{ij}=1$, $\zeta ^{\ast
}=0$, $ \gamma =1$, and Eqs.\ (\ref{2.9})--(\ref{2.11}) become
\begin{equation}
D^{\ast }=\frac{3}{8}\frac{1+\delta }{1-\mu _{12}(1-\delta
)},\quad D_{p}^{\ast }=x_{1}\frac{(1-\mu )}{1+\mu \delta }D^{\ast
}, \quad D^{\prime }{}^{\ast }=0. \label{2.11e}
\end{equation}
These coincide with known results obtained for elastic collisions
in the first Sonine approximation. \cite{CC70} Recently, it has
been shown that the estimate given by the first Sonine
approximation for the diffusion coefficient $D$ (in the very
dilute concentration limit $x_{1}\rightarrow 0$) compares quite
well with Monte Carlo simulations, \cite{GM04} except for the
cases in which the gas particles are much heavier and/or much
larger than impurities. For these extreme cases, the second Sonine
approximation to $D$ improves the accuracy of the kinetic theory
results. \cite{GM04}

\subsection{Heat flux}
\label{subsec2.2}

The heat flux requires going up to the second Sonine
approximation. The transport coefficients $D^{\prime \prime }$,
$L$, and $\lambda $ appearing in the heat flux (\ref{2.2}) are
given by \cite{GD02}

\begin{equation}
D^{\prime \prime }=-\frac{5}{2}\frac{n}{(m_{1}+m_{2})\nu
_{0}}\left[ \frac{ x_{1}\gamma _{1}^{3}}{\mu _{12}}d_{1}^{\prime
\prime }+\frac{x_{2}\gamma _{2}^{3}}{\mu _{21}}d_{2}^{\prime
\prime }-\left( \frac{\gamma _{1}}{\mu _{12}}-\frac{\gamma
_{2}}{\mu _{21}}\right) D^{\ast }\right] ,  \label{2.12}
\end{equation}
\begin{equation}
L=-\frac{5}{2}\frac{T}{(m_{1}+m_{2})\nu _{0}}\left[
\frac{x_{1}\gamma _{1}^{3}}{\mu _{12}}\ell _{1}+\frac{x_{2}\gamma
_{2}^{3}}{\mu _{21}}\ell _{2}-\left( \frac{\gamma _{1}}{\mu
_{12}}-\frac{\gamma _{2}}{\mu _{21}} \right) D_{p}^{\ast }\right]
,  \label{2.13}
\end{equation}
\begin{equation}
\lambda =-\frac{5}{2}\frac{nT}{(m_{1}+m_{2})\nu _{0}}\left[ \frac{
x_{1}\gamma _{1}^{3}}{\mu _{12}}\lambda _{1}+\frac{x_{2}\gamma
_{2}^{3}}{\mu_{21}}\lambda _{2}-\left( \frac{\gamma _{1}}{\mu
_{12}}-\frac{\gamma _{2}}{ \mu _{21}}\right) D^{\prime }{}^{\ast
}\right] ,  \label{2.14}
\end{equation}
where the expressions for the (dimensionless) Sonine coefficients
$\{d_{i}^{\prime \prime },\ell _{i},\lambda _{i}\}$ are displayed
in Appendix \ref{appA}. In Eqs.\ (\ref{2.12})--(\ref{2.14}) it is
understood that the coefficients $D^{\ast }$, $D_{p}^{\ast }$, and
 $D^{\prime \ast }$ are given by
 Eqs.\ (\ref{2.9})--(\ref{2.11}), respectively (first
Sonine approximation). As expected, our results show that
$D^{\prime \prime }$ is antisymmetric with respect to the change
$1\leftrightarrow 2$ while $L$ and $ \lambda $ are symmetric.
Consequently, in the case of mechanically equivalent particles
($m_{1}=m_{2}\equiv m$, $\sigma _{1}=\sigma _{2}\equiv \sigma$,
$\alpha _{ij}\equiv \alpha $), the coefficient $D^{\prime \prime}$
vanishes.

An equivalent representation is given in terms of the heat flow
$\mathbf{J}_{q}$ defined as
\begin{equation}
\mathbf{J}_{q}\equiv
\mathbf{q}-\frac{5}{2}T\sum_{i}\frac{\mathbf{j}_{i}}{
m_{i}}=\mathbf{q}-\frac{5}{2}T\frac{m_{2}-m_{1}}{m_{1}m_{2}}\mathbf{j}_{1},
\label{2.15}
\end{equation}
where in the second equality use has been made of the requirement
$\mathbf{j} _{1}=-\mathbf{j}_{2}$. The difference between
$\mathbf{q}$ and $\mathbf{J} _{q}$ is a heat flow due to
diffusion. In addition, for elastic collisions, $ \mathbf{J}_{q}$
is the flux conjugate to the temperature gradient in the form of
the entropy production where the contribution coming from the mass
flux couples only to the gradient of the chemical potentials. The
thermal conductivity in a mixture is generally measured in the
absence of diffusion, i.e., when $\mathbf{j}_{1}=\mathbf{0}$. To
identify this coefficient, we have to express $\mathbf{J}_{q}$ in
terms of $\mathbf{j}_{1}$, $\nabla T$, and $\nabla p$. The
corresponding coefficient of $\nabla T$ defines the thermal
conductivity. \cite{GM84} According to Eq.\ (\ref{2.1}), the
gradient of mole fraction $\nabla x_{1}$ is
\begin{equation}
\nabla x_{1}=-\frac{\rho }{m_{1}m_{2}nD}\mathbf{j}_{1}-\frac{\rho
^{2}}{ m_{1}m_{2}p}\frac{D^{\prime }}{D}\nabla T-\frac{\rho
^{2}}{m_{1}m_{2}np} \frac{D_{p}}{D}\nabla p.  \label{2.16}
\end{equation}
The expression of $\mathbf{J}_{q}$ is obtained by substituting
Eq.\ (\ref {2.2}) into Eq.\ (\ref{2.15}) and eliminating $\nabla
x_{1}$ by using the identity (\ref{2.16}). Thus, the heat flow is
given by
\begin{equation}
\mathbf{J}_{q}=-\kappa \nabla T+\frac{\rho
p}{m_{1}m_{2}n^{2}}\kappa_{\text{T}}\mathbf{j}_{1}-L_{p}\nabla p,
\label{2.17}
\end{equation}
where
\begin{equation}
\kappa =\lambda -\frac{\rho ^{2}T^{2}}{m_{1}m_{2}p}\frac{D^{\prime \prime
}D^{\prime }}{D},  \label{2.18}
\end{equation}
\begin{equation}
\kappa _{\text{T}}=T\frac{D^{\prime \prime
}}{D}-\frac{5}{2}\frac{n}{\rho } (m_{2}-m_{1}),  \label{2.19}
\end{equation}
\begin{equation}
L_{p}=L-\frac{\rho ^{2}T}{n^{2}m_{1}m_{2}}\frac{D^{\prime \prime }D_{p}}{D}.
\label{2.20}
\end{equation}
As in the elastic case, the coefficient $\kappa $ is the thermal
conductivity while $\kappa _{\text{T}}$ is called the thermal-diffusion
factor or Dufour coefficient. There is a \emph{new} contribution
proportional to $\nabla p$ not present in the elastic case that defines the
transport coefficient $L_{p}$.

For elastic collisions, $L_{p}=0$ \cite{comment} and the
expressions derived here for $\kappa $ and $\kappa _{\text{T}}$
coincide with those obtained for a gas-mixture of elastic hard
spheres. \cite{CC70} Furthermore, in the case of mechanically
equivalent particles, the Dufour coefficient $\kappa _{\text{T}}$
vanishes as expected and the heat flow (\ref {2.17}) can be
written as
\begin{equation}
\mathbf{J}_{q}=-\overline{\kappa }\nabla T-\overline{\mu }\nabla n,
\label{2.21}
\end{equation}
where
\begin{equation}
\overline{\kappa }=\kappa +nL_{p}=\frac{25}{32}\left( \frac{mT}{\pi }\right)
^{1/2}\sigma ^{-2}\frac{1+c}{\nu _{\kappa }-2\zeta ^{\ast }},  \label{2.22}
\end{equation}
\begin{equation}
\overline{\mu }=TL_{p}=\frac{75}{32}\frac{T}{n}\left(
\frac{mT}{\pi }\right) ^{1/2}\sigma ^{-2}\zeta ^{\ast }\left(
\frac{2}{3}\frac{1+c}{\nu _{\kappa }-2\zeta ^{\ast
}}+\frac{1}{3}\frac{c}{\zeta ^{\ast }}\right) \left(2\nu _{\kappa
}-3\zeta ^{\ast }\right) ^{-1},  \label{2.23}
\end{equation}
\begin{equation}
\nu _{\kappa }=\frac{1}{3}(1+\alpha )\left[
1+\frac{33}{16}(1-\alpha )+\frac{ 19-3\alpha }{1024}c\right] ,
\label{2.24}
\end{equation}
\begin{equation}
c=\frac{32(1-\alpha )(1-2\alpha ^{2})}{81-17\alpha +30\alpha
^{2}(1-\alpha )} ,\quad \zeta ^{\ast }=\frac{5}{12}(1-\alpha
^{2})\left( 1+\frac{3}{32} c\right) .  \label{2.25}
\end{equation}
Note that in writing Eq.\ (\ref{2.21}) use has been made of the
relation $ \nabla p=n\nabla T+T\nabla n$. Equations
(\ref{2.21})--(\ref{2.25}) are the same as those obtained for the
one component granular gas. \cite{BDKS98} This confirms the
relevant known limiting cases for the granular mixture results
described here.

\section{Transport coefficients}
\label{sec3}

The transport coefficients depend on many parameters: $\left\{
x_{1},T,m_{1}/m_{2},\sigma _{1}/\sigma _{2}, \alpha _{11}, \alpha
_{22}, \alpha _{12}\right\} $. This complexity exists in the
elastic limit as well, so the primary new feature is the
dependence on the coefficients of restitution  $ \alpha _{ij}$
being different from unity. To illustrate the differences between
granular and normal gases the transport coefficients are
normalized to their values in the elastic limit. Then, the
dependence on the overall temperature scales out. Also, only the
simplest case of a common coefficient of restitution ($\alpha
_{11}=\alpha _{22}=\alpha _{12}\equiv \alpha $) and common size
$\omega \equiv \sigma _{1}/\sigma _{2}=1$ is considered. This
reduces the parameter set to three quantities:
$\{m_{1}/m_{2},x_{1},\alpha \}$.

\begin{figure}[tbp]
\begin{center}
\resizebox{7.5cm}{!}{\includegraphics{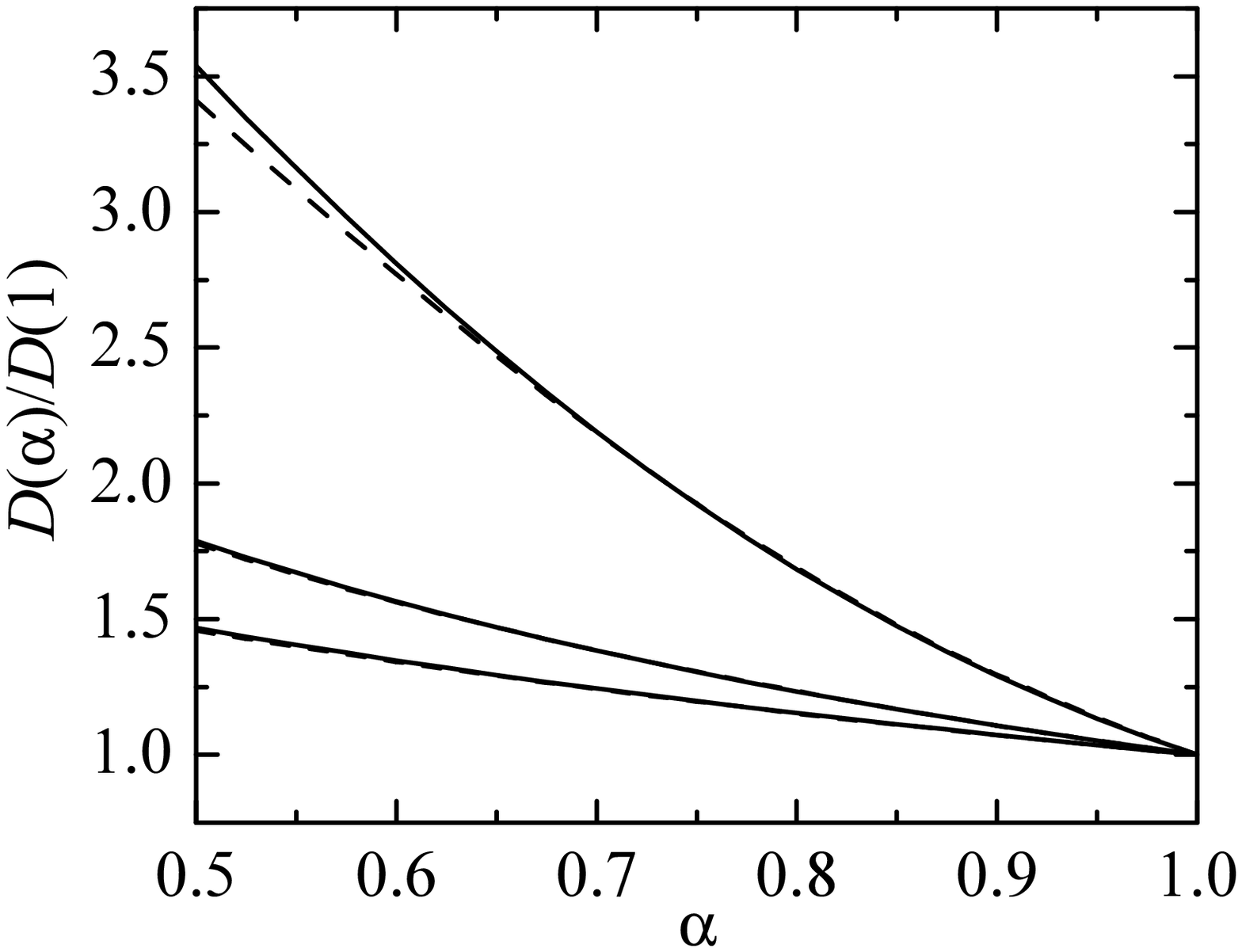}}
\end{center}
\caption{Plot of the reduced mutual diffusion coefficient
$D(\protect\alpha )/D(1)$ as a function of the coefficient of
restitution $\protect\alpha$ for $x_1=0.2$, $\protect\omega=1$ and
$\protect\mu=0.5$ (a), $\protect\mu=1$ (b), and $\protect\mu=4$
(c). The dashed lines correspond to the approximation
$c_1=c_2=0$.} \label{fig1}
\end{figure}

\begin{figure}[tbp]
\begin{center}
\resizebox{7.5cm}{!}{\includegraphics{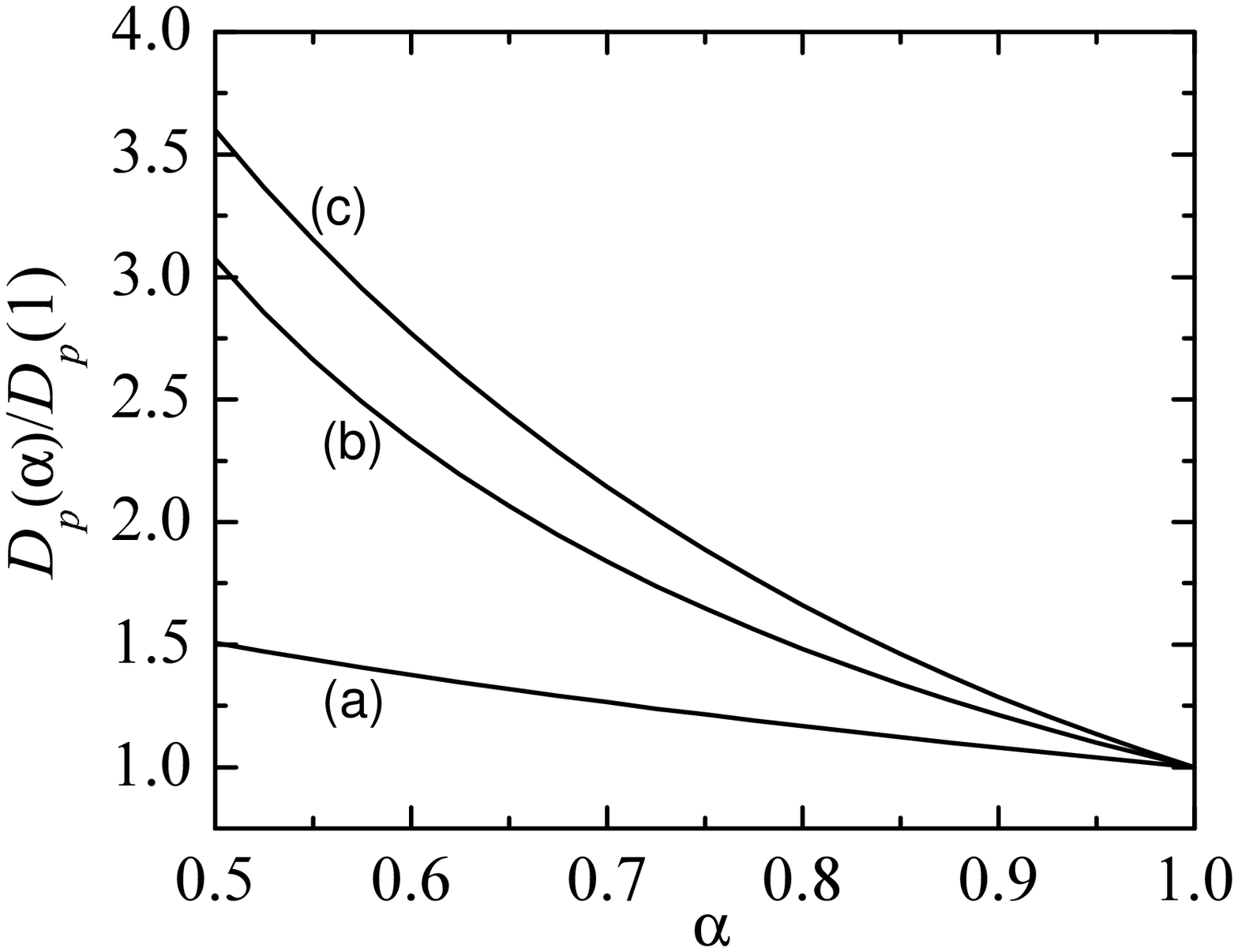}}
\end{center}
\caption{Plot of the reduced pressure diffusion coefficient
$D_p(\protect \alpha)/D_{p}(1)$ as a function of the coefficient
of restitution $\protect \alpha$ for $x_1=0.2$, $\protect\omega=1$
and $\protect\mu=0.5$ (a), $ \protect\mu=2$ (b), and
$\protect\mu=4$ (c).} \label{fig2}
\end{figure}

\begin{figure}[tbp]
\begin{center}
\resizebox{7.5cm}{!}{\includegraphics{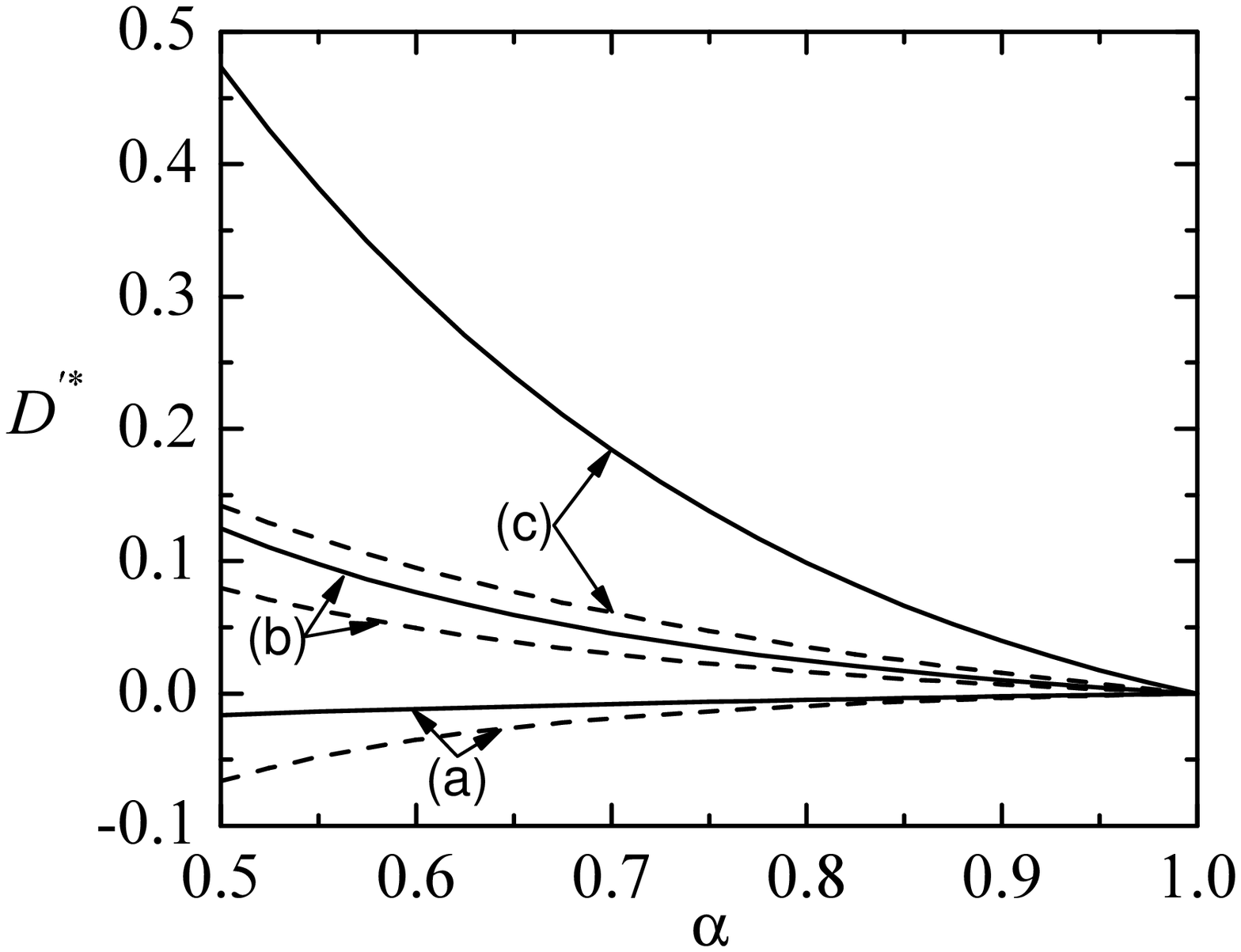}}
\end{center}
\caption{Plot of the reduced thermal diffusion coefficient
$D^{^{\prime }\ast }(\protect\alpha )$ as a function of the
coefficient of restitution $ \protect\alpha $ for $x_{1}=0.2$,
$\protect\omega =1$ and $\protect\mu =0.5$ (a), $\protect\mu =2$
(b), and $\protect\mu =4$ (c). The dashed lines correspond to
$\protect\gamma =1$.} \label{fig3}
\end{figure}

\begin{figure}[tbp]
\begin{center}
\resizebox{7.5cm}{!}{\includegraphics{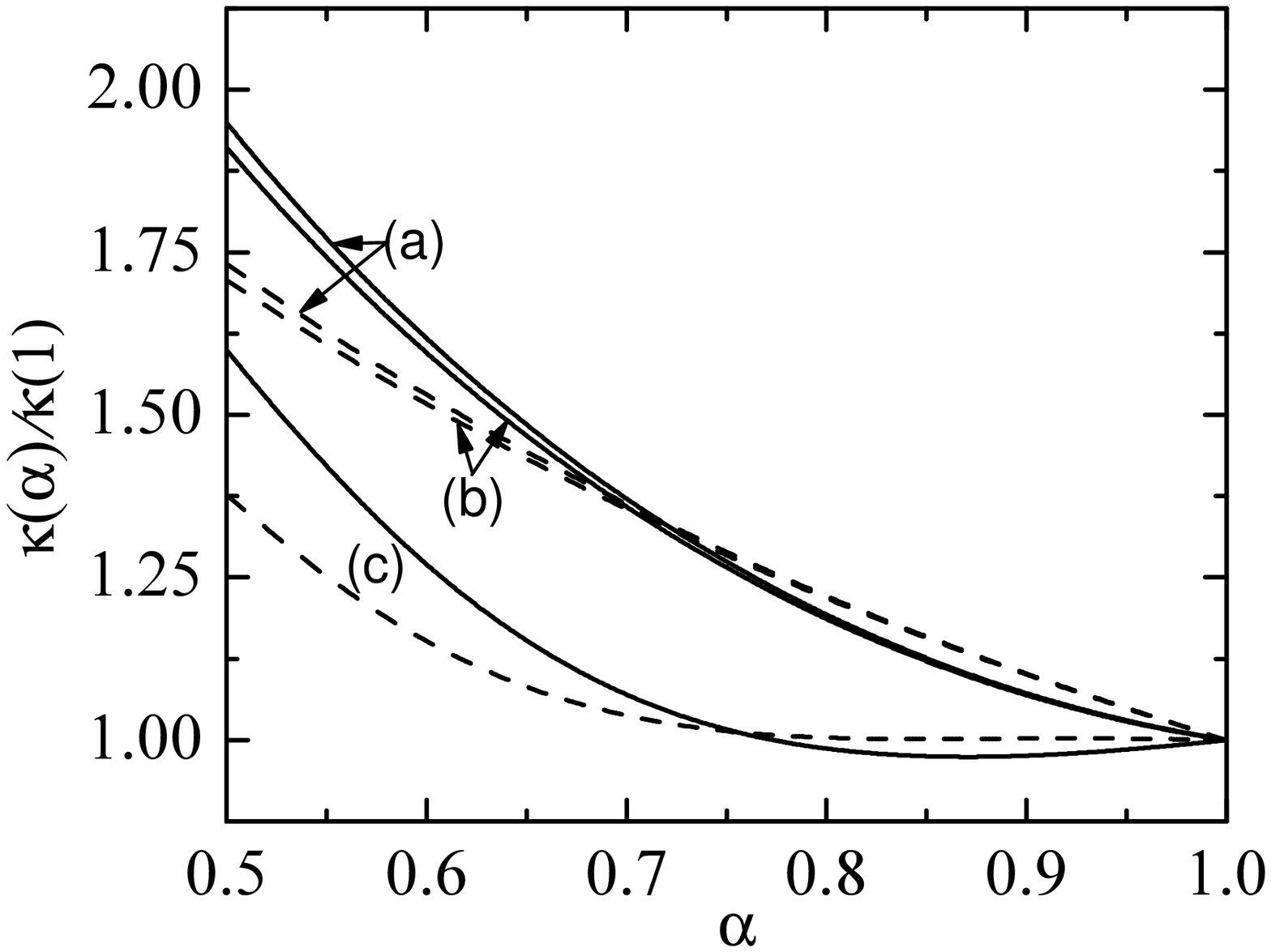}}
\end{center}
\caption{Plot of the reduced thermal conductivity coefficient
$\protect\kappa (\protect\alpha)/\protect\kappa(1)$ as a function
of the coefficient of restitution $\protect\alpha$ for $x_1=0.2$,
$\protect\omega=1$ and $\protect \mu=0.5$ (a), $\protect\mu=1$
(b), and $\protect\mu=4$ (c). The dashed lines correspond to the
approximation $c_1=c_2=0$.} \label{fig4}
\end{figure}

\begin{figure}[tbp]
\begin{center}
\resizebox{7.5cm}{!}{\includegraphics{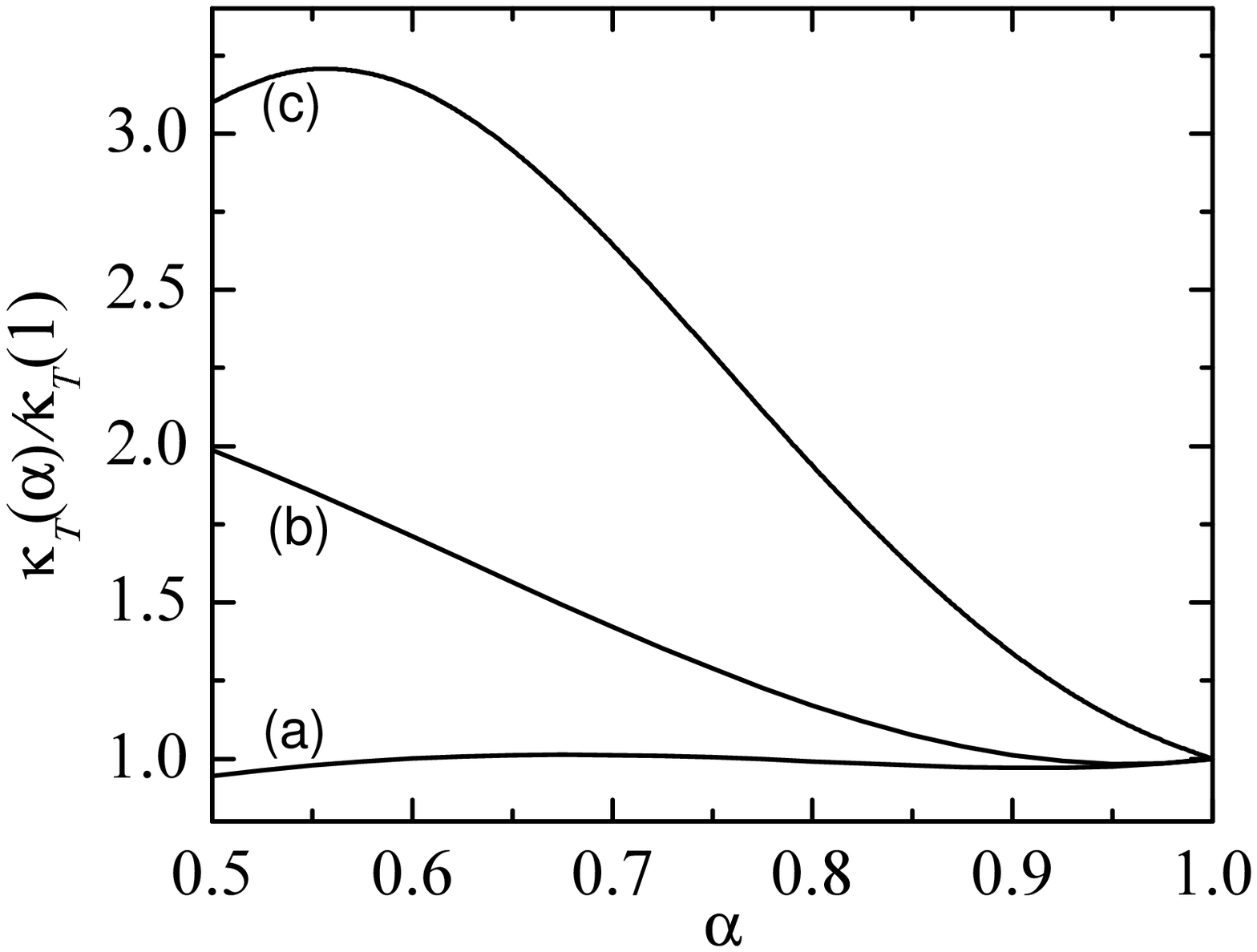}}
\end{center}
\caption{Plot of the reduced Dufour coefficient
$\protect\kappa_T(\protect \alpha)/\protect\kappa_{T}(1)$ as a
function of the coefficient of restitution $\protect\alpha$ for
$x_1=0.2$, $\protect\omega=1$ and $\protect \mu=0.5$ (a),
$\protect\mu=2$ (b), and $\protect\mu=4$ (c).} \label{fig5}
\end{figure}

\begin{figure}[tbp]
\begin{center}
\resizebox{7.5cm}{!}{\includegraphics{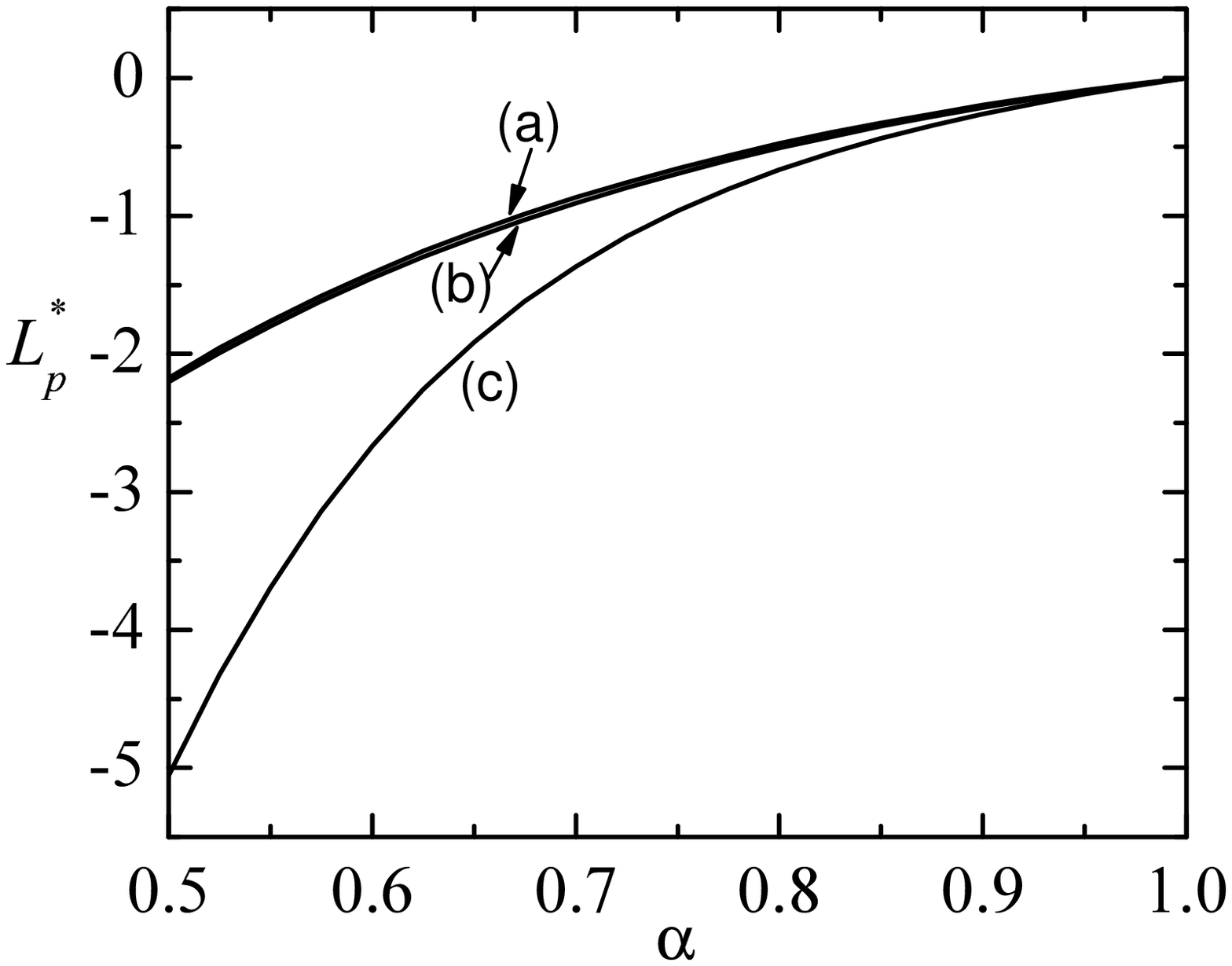}}
\end{center}
\caption{Plot of the reduced coefficient $L_p^*(\protect\alpha)$
as a function of the coefficient of restitution $\protect\alpha$
for $x_1=0.2$, $ \protect\omega=1$ and $\protect\mu=0.5$ (a),
$\protect\mu=1$ (b), and $ \protect\mu=4$ (c).} \label{fig6}
\end{figure}

In Figs.\ \ref{fig1}--\ref{fig6}, we plot the above transport
coefficients as functions of the coefficient of restitution
$\alpha $ for $x_{1}=0.2$, and several values of the mass ratio
$\mu$. It is understood that all coefficients have been reduced
with respect to their elastic values, except in the cases of
$D^{\prime }$ and $L_{p}$ since both coefficients vanish for
elastic collisions. In these latter two cases, we have considered
the reduced coefficients $D^{\prime }{}^{\ast }$ defined by Eq.\
(\ref{2.9}) and $ L_{p}^{\ast }=-\left[ (5/2)T\nu
_{0}/(m_{1}+m_{2})\right]^{-1}L_{p}$. Figure \ref{fig1} shows the
mutual diffusion coefficient as a function of $ \alpha $ for three
mass ratios $\mu=0.5,1,$ and $4$. There is a monotonic increase of
the coefficient with decreasing $\alpha $ in all cases. Moreover,
that effect increases as the mass of the dilute species increases.
This is consistent with an observed singular behavior in the
extreme case of tracer diffusion for a massive particle.
\cite{SD01} The velocity distributions in a granular gas are no
longer Maxwellian, \cite{GD99} and the difference is measured by
the coefficients $c_{i}$ that appear in the expressions for the
transport coefficients. \cite{GD02} The dashed curves in Fig.\
\ref{fig1} correspond to $c_{1}=c_{2}\rightarrow 0$, the Maxwell
limit. In this case it is seen that the distortion of the
Maxwellian is not very important for this transport coefficient
(see however, discussion of Fig.\ \ref{fig4} below). Figure
\ref{fig2} shows that the pressure diffusion coefficient has a
very similar behavior. The thermal diffusion coefficient vanishes
in the elastic limit and remains small and slightly negative when
the dilute species has small mass ratio, as illustrated in Fig.\
\ref{fig3}. However as the mass ratio becomes large it becomes
large and positive for strong dissipation. The effect of different
species temperatures is also shown on this graph. The dashed
curves correspond to setting $\gamma =T_{1}/T_{2}\rightarrow 1$.
This is seen to yield large errors, particularly in the case of
large mass ratio (temperature differences are greater for
mechanically different particles), indicating the real
quantitative effect of two different species temperatures in
granular gases.

The thermal conductivity is shown in Fig.\  \ref{fig4}, with the
same monotonic increase with increasing dissipation. The dashed
lines ($c_i=0$) indicate a significant effect of the distortion of
the reference distribution function from its Maxwellian form.
Presumably, this is due to the fact that the thermal conductivity
$\kappa$ depends on a higher velocity moment than the mutual
diffusion coefficient $D$ and is more sensitive to the larger
distortions at higher velocities. We also observe that there is
little mass dependence when the dilute species is lighter.
However, when the dilute species is more massive there is a
significant decrease in the thermal conductivity, opposite to the
case of diffusion. In contrast, the Dufour coefficient does have a
dependence on the mass ratio more like diffusion (Fig.\
\ref{fig5}). Finally, the coefficient $L_{p}^{\ast }$ is
illustrated in Figure \ref{fig6}. Again there is weak dependence
on the mass ratio until it becomes larger for the dilute species.

In summary, the mass and heat flux transport coefficients for a
granular mixture differ significantly from those for a normal gas
mixture even at moderate dissipation. In most cases (thermal
conductivity is an exception) the differences increase with
decreasing $\alpha $, depend weakly on the mass ratio when the
dilute species ($x_1/x_2<1$) is lighter than the excess species
($m_{1}/m_{2}\leq 1$) but increase significantly in the opposite
case ($m_{1}/m_{2}>1$).


\section{Onsager's reciprocal relations}
\label{sec4}

In the usual language of the linear irreversible thermodynamics
for ordinary fluids, \cite{GM84} the constitutive equations for
the mass flux (\ref{2.1}) and heat flow (\ref{2.17}) is written
\begin{equation}
\mathbf{j}_{i}=-\sum_{i}L_{ij}\left( \frac{\nabla \mu
_{j}}{T}\right) _{T}-L_{iq}\frac{\nabla T}{T^{2}}-C_{p}\nabla p,
\label{4.1}
\end{equation}
\begin{equation}
\mathbf{J}_{q}=-L_{qq}\nabla T-\sum_{i}L_{qi}\left( \frac{\nabla
\mu _{i}}{T} \right) _{T}-C_{p}^{\prime }\nabla p,  \label{4.2}
\end{equation}
where
\begin{equation}
\left( \frac{\nabla \mu _{i}}{T}\right) _{T}=\frac{1}{m_{i}}\nabla
\ln (x_{i}p),  \label{4.3}
\end{equation}
$\mu _{i}$ being the chemical potential per unit mass. Here, the
coefficients $L_{ij}$ are the so-called Onsager phenomenological
coefficients. For \emph{normal} fluids, Onsager showed \cite{GM84}
that time reversal invariance of the underlying microscopic
equations of motion implies important restrictions on the above
set of transport coefficients
\begin{equation}
L_{ij}=L_{ji},\quad L_{iq}=L_{qi},\quad C_{p}=C_{p}^{\prime }=0.
\label{4.4}
\end{equation}
The first two symmetries are called reciprocal relations as they
relate transport coefficients for different processes. The last
two are statements that the pressure gradient does not appear in
any of the fluxes even though it is admitted by symmetry. Even for
a one component fluid, Onsager's theorem is significant as it
leads to Fourier's law for the heat flow rather than (\ref{2.21}),
i.e. $\overline{\mu }=0$. Since there is no time reversal symmetry
for granular fluids, Eqs. (\ref{4.4}) cannot be expected to apply.
However, since explicit expressions for all transport coefficients
are at hand, the quantitative extent of the violation can be
explored.

To make connection with the previous sections it is first
necessary to transform Eqs.\ (\ref{4.1})--(\ref{4.3}) to the
variables $x_{1},$ $p,$ $T.$ Since $\nabla x_{1}=-\nabla x_{2}$,
Eq.\ (\ref{4.3}) implies
\begin{equation}
\frac{(\nabla \mu _{1})_{T}-(\nabla \mu _{2})_{T}}{T}=\frac{n\rho
}{\rho _{1}\rho _{2}}\left[ \nabla x_{1}+\frac{n_{1}n_{2}}{n\rho }
(m_{2}-m_{1})\nabla \ln p\right] .  \label{4.5}
\end{equation}
The coefficients $L_{ij}$ then can be easily obtained in terms of
those of the previous sections. The result is
\begin{equation}
L_{11}=-L_{12}=-L_{21}=\frac{m_{1}m_{2}\rho _{1}\rho _{2}}{\rho
^{2}}D,\quad L_{1q}=\rho TD^{\prime },  \label{4.6}
\end{equation}
\begin{equation}
L_{q1}=-L_{q2}=\frac{T^{2}\rho _{1}\rho _{2}}{n\rho }D^{\prime
\prime }- \frac{5}{2}\frac{T\rho _{1}\rho _{2}}{\rho
^{2}}(m_{2}-m_{1})D,\quad L_{qq}=\lambda -\frac{5}{2}\rho
\frac{m_{2}-m_{1}}{m_{1}m_{2}}D^{\prime }, \label{4.7}
\end{equation}
\begin{equation}
C_{p}\equiv \frac{\rho}{p} D_{p}-\frac{\rho _{1}\rho _{2}}{p\rho
^{2}}(m_{2}-m_{1})D, \label{4.8}
\end{equation}
\begin{equation}
C_{p}^{\prime }\equiv
L-\frac{5}{2}\frac{T}{p}\frac{m_{2}-m_{1}}{m_{1}m_{2}}C_{p}-
\frac{n_{1}n_{2}}{np\rho }T^{2}(m_2-m_1)D^{\prime \prime}.
\label{4.9}
\end{equation}
The Onsager's relation $L_{12}=L_{21}$ is already evident since
the diffusion coefficient $D$ is symmetric under the change
$1\leftrightarrow 2$ , as discussed following Eq.\ (\ref{2.11a})

Imposing Onsager's relation $L_{1q}=L_{q1}$ yields
\begin{equation}
D^{\prime \prime }=\frac{5}{2}\frac{n}{T\rho
}(m_{2}-m_{1})D+\frac{n\rho ^{2} }{T\rho _{1}\rho _{2}}D^{\prime
},  \label{4.10}
\end{equation}
while the condition $C_{p}=C_{p}^{\prime }=0$ leads to the
following additional requirements
\begin{equation}
D_{p}=\frac{\rho _{1}\rho _{2}}{\rho ^{3}}(m_{2}-m_{1})D,
\label{4.11}
\end{equation}
\begin{equation}
\frac{5}{2}\frac{T}{p}(m_{1}-m_{2})\left[ \frac{n_{1}n_{2}}{\rho
^{2}}(m_{2}-m_{1})D- \frac{\rho }{m_{1}m_{2}}D_{p}\right]
=pL-\frac{n_{1}n_{2}}{n\rho } T^{2}(m_{2}-m_{1})D^{\prime \prime
}.  \label{4.12}
\end{equation}
Since the relations (\ref{4.10})--(\ref{4.12}) involve transport
coefficients that have been determined in the first Sonine
approximation, we restrict here our discussion to this level of
approximation. In this case, $ d_{i}^{\prime \prime }=\ell
_{i}=\lambda _{i}=0$ so that Onsager's theorem, Eqs.\
(\ref{4.10})--(\ref{4.12}), gives the conditions
\begin{equation}
P(\alpha _{ij})\equiv \left[ \gamma _{1}-1+\mu (1-\gamma
_{2})\right] \frac{ D^{\ast }}{\mu }+\frac{1}{5}\frac{(1+\delta
)(1+\mu \delta )}{\mu \delta } \frac{\zeta ^{\ast }}{\nu ^{\ast
}}D_{p}^{\ast }=0,  \label{4.13}
\end{equation}
\begin{equation}
Q(\alpha _{ij})\equiv D_{p}^{\ast }-x_{1}\frac{(1-\mu )}{(1+\mu
\delta )} D^{\ast }=0,  \label{4.14}
\end{equation}
\begin{equation}
R(\alpha _{ij})\equiv \frac{1+\mu }{\mu }\left( \gamma _{1}-\mu
\gamma _{2}\right) Q(\alpha _{ij})=0.  \label{4.15}
\end{equation}

\begin{figure}[tbp]
\begin{center}
\resizebox{7.5cm}{!}{\includegraphics{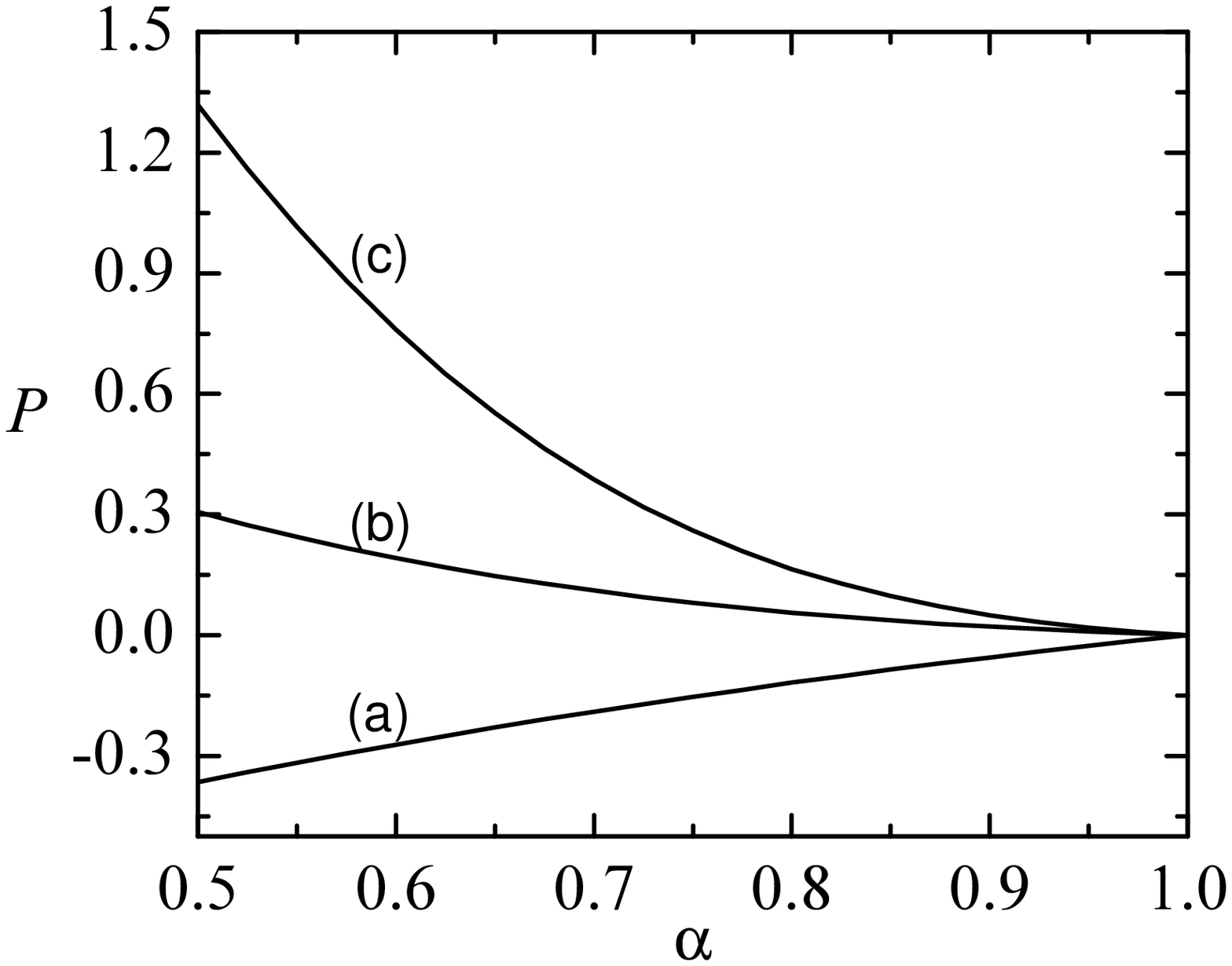}}
\end{center}
\caption{Plot of $P(\protect\alpha)$ as a function of
$\protect\alpha$ for $ x_1=0.2$, $\protect\omega=1$ and
$\protect\mu=0.5$ (a), $\protect\mu=2$ (b), and $\protect\mu=4$
(c).} \label{fig7}
\end{figure}

\begin{figure}[tbp]
\begin{center}
\resizebox{7.5cm}{!}{\includegraphics{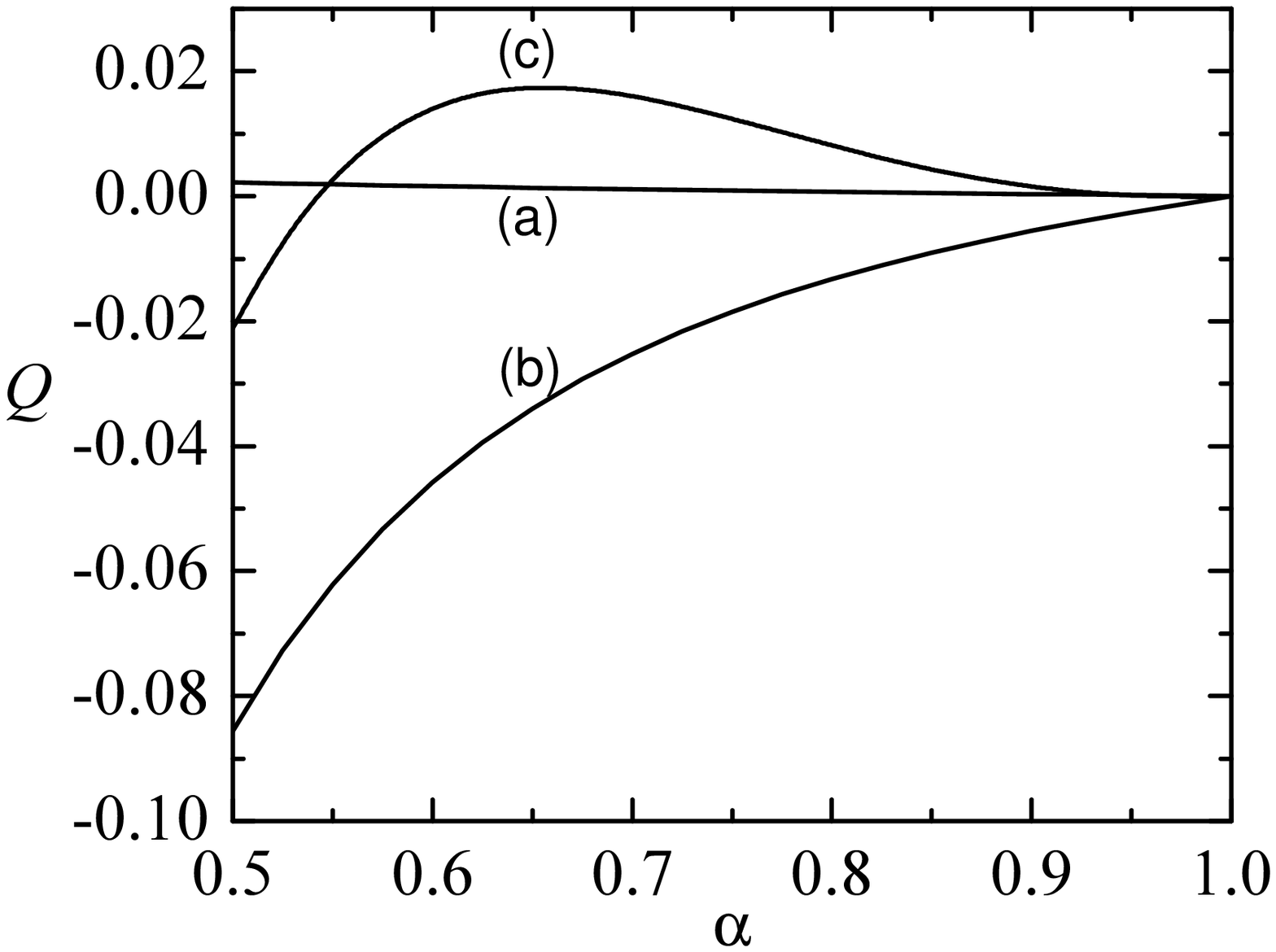}}
\end{center}
\caption{Plot of $Q(\protect\alpha)$ as a function of
$\protect\alpha$ for $ x_1=0.2$, $\protect\omega=1$ and
$\protect\mu=0.5$ (a), $\protect\mu=2$ (b), and $\protect\mu=4$
(c).} \label{fig8}
\end{figure}

\begin{figure}[tbp]
\begin{center}
\resizebox{7.5cm}{!}{\includegraphics{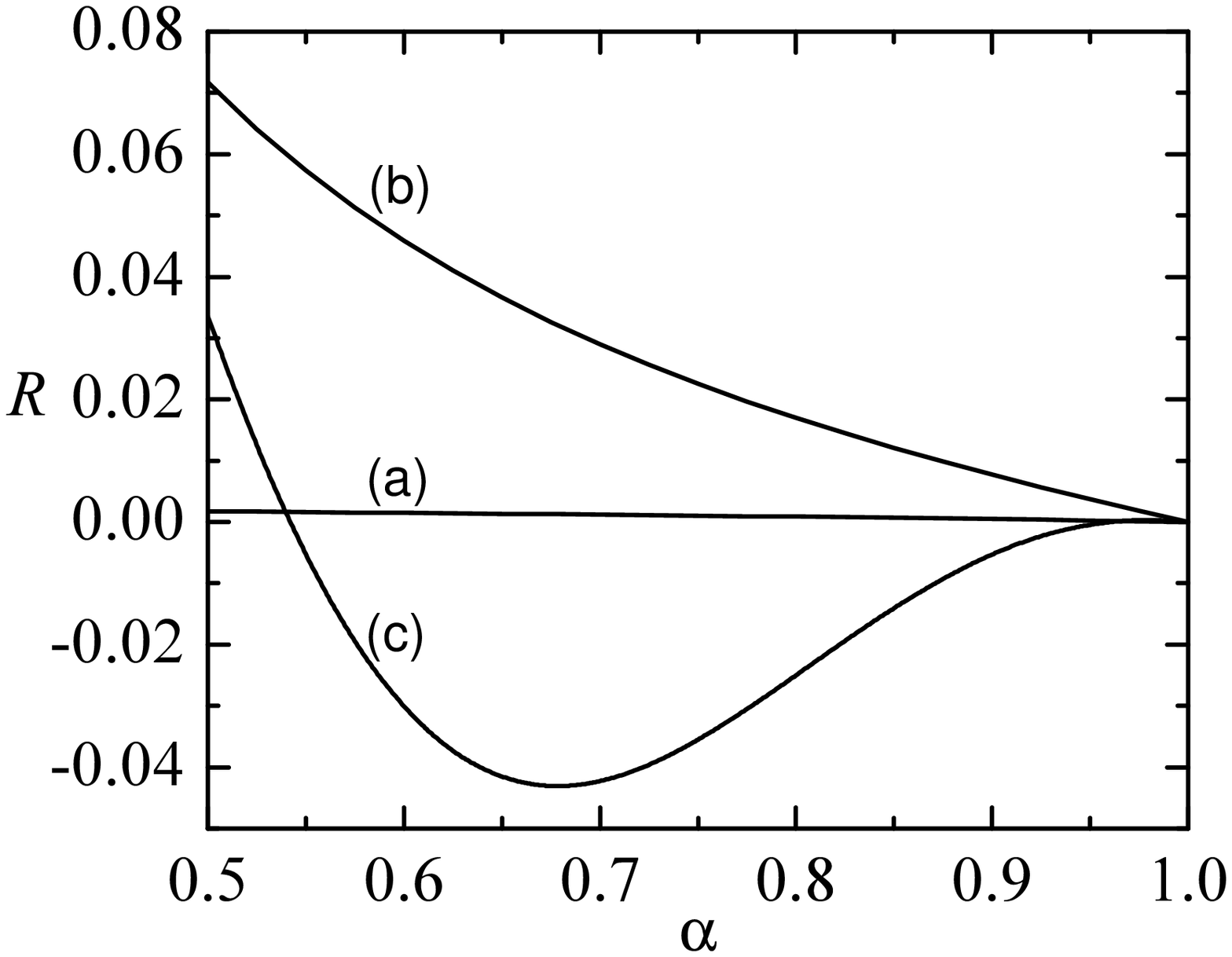}}
\end{center}
\caption{Plot of $R(\protect\alpha)$ as a function of
$\protect\alpha$ for $ x_1=0.2$, $\protect\omega=1$ and
$\protect\mu=0.5$ (a), $\protect\mu=2$ (b), and $\protect\mu=4$
(c).} \label{fig9}
\end{figure}

In the elastic limit, the reduced coefficients $D_{p}^{\ast }$ and
$D^{\ast } $ are given by Eq.\ (\ref{2.11}) and these conditions
are verified. Also, for mechanically equivalent particles with
arbitrary $\alpha $, $\gamma _{i}=1$ and $D_{p}^{\ast }=0$ so that
$P(\alpha )=Q(\alpha )=R(\alpha )=0$. Nevertheless, beyond these
limit cases, Onsager's relations do not apply (as expected). At
this macroscopic level the origin of this failure is due to the
cooling of the reference state as well as the occurrence of
different kinetic temperatures for both species. Figures
\ref{fig7}, \ref{fig8}, and \ref{fig9} show the dependence of the
quantities $P$, $Q$, and $R$, respectively, on the (common)
coefficient of restitution $\alpha_{ij}\equiv \alpha$ for mass
ratios $\mu =0.5,2,$ and $4$.

\section{Linearized hydrodynamic equations and stability}
\label{sec5}

In contrast to normal fluids, the Navier--Stokes hydrodynamic
equations\ (\ref {2.4})--(\ref{2.7}) have non-trivial solutions
even for spatially homogeneous states,
\begin{equation}
\label{5.0}
\partial _{t}x_{1H}=0=\partial _{t}u_{H\ell },
\end{equation}
\begin{equation}
\left[ \partial _{t}+\zeta \left( x_{1H},T_{H},p_{H}\right)
\right] T_{H}=0, \hspace{0.3in}\left[ \partial _{t}+\zeta \left(
x_{1H},T_{H},p_{H}\right) \right] p_{H}=0.  \label{5.1}
\end{equation}
where the subscript $H$ denotes the homogeneous state. Since the
dependence of the cooling rate $\zeta \left(
x_{1H},T_{H},p_{H}\right)$ on $x_{1H},T_{H},p_{H}$ is known (see
Appendix A), these first order nonlinear equations can be solved
for the time dependence of the homogeneous state. The result is
the familiar Haff's law cooling law for $T(t)$ at constant
density. \cite{BP04}. As discussed above, each species temperature
also has the same time dependence but each with a different value
\cite{GD99} In this section, the hydrodynamics for small initial
spatial perturbations of this homogeneous cooling state (HCS) is
discussed. For normal fluids such perturbations decay in time
according to the hydrodynamic modes of diffusion (shear, thermal,
mass) and damped sound propagation.  The analysis is for fixed
coefficients of restitution different from unity in the long
wavelength limit. It will be seen here that the corresponding
modes for a granular fluid are then quite different from those for
a normal fluid. An alternative analysis with fixed long wavelength
and coefficient of restitution  approaching unity leads to usual
normal fluid modes. Thus, the nature of hydrodynamic modes is non
uniform with respect to the inelasticity and the wavelength of the
perturbation.

Let $\delta y_{\alpha }(\mathbf{r},t)=y_{\alpha
}(\mathbf{r},t)-y_{H\alpha }(t)$ denote the deviation of
$\{x_{1},\mathbf{u},T,p\}$ from their values in the HCS. If the
initial spatial perturbation is sufficiently small, then for some
initial time interval these deviations will remain small and the
hydrodynamic equations (\ref{2.4})--(\ref{2.7}) can be linearized
with respect to $\delta y_{\alpha }(\mathbf{r},t)$. This leads to
partial differential equations with coefficients that are
independent of space but which depend on time since the HCS is
cooling. As in the one component case, \cite{BDKS98,G05} this time
dependence can be eliminated through a change in the time and
space variables, and a scaling of the hydrodynamic fields.  We
introduce the following dimensionless space and time variables:
\begin{equation}
\tau =\int_{0}^{t}dt^{\prime }\nu_{H}(t^{\prime }),\quad
\mathbf{r}^{\prime }= \mathbf{r/}\ell _{H},  \label{5.2}
\end{equation}
where $v_{0H}=\sqrt{2T_{H}(m_{1}+m_{2})/m_{1}m_{2}}$ is the
thermal velocity introduced above, $\ell_{H}=1/\sqrt{\pi }
n_{H}\sigma _{12}^{2}$ is an effective mean free path, and
$\nu_{H}(t)=v_{0H}(t)/\ell _{H}$ is the effective collision
frequency. The dimensionless time scale is therefore an average
number of collisions up to the time $t$. A set of Fourier
transformed dimensionless variables are then defined by
\begin{equation}
\delta y_{\mathbf{k}\alpha }(\tau )=\int
d\mathbf{r}^{\prime}\;e^{-i\mathbf{k} \cdot
\mathbf{r}^{\prime}}\delta y_{\alpha }(\mathbf{r}^{\prime },
\tau), \label{5.3}
\end{equation}
\begin{equation}
\rho_{\mathbf{k}}(\tau )=\frac{\delta x_{1\mathbf{k}}(\tau
)}{x_{1H}},\quad \mathbf{w}_{\mathbf{k}}(\tau )=\frac{\delta
\mathbf{u}_{\mathbf{k}}(\tau )}{ v_{0H}(\tau )},  \label{5.4}
\end{equation}
\begin{equation}
\theta _{\mathbf{k}}(\tau )=\frac{\delta T_{\mathbf{k}}(\tau
)}{T_{H}(\tau )} ,\quad \Pi _{\mathbf{k}}(\tau )=\frac{\delta
p_{\mathbf{k}}(\tau )}{ p_{H}(\tau )}.  \label{5.5}
\end{equation}
In terms of these variables the linearized hydrodynamic equations
for the set $\{\rho _{\mathbf{k}},\mathbf{w}_{\mathbf{k}},\theta
_{\mathbf{k}},\Pi _{ \mathbf{k}}\}$ separate into a degenerate
pair of equations for the two transverse velocity components
$w_{\mathbf{k}\perp }$ (orthogonal to $ \mathbf{k}$)
\begin{equation}
\left( \frac{\partial }{\partial \tau }-\frac{\zeta ^{\ast
}}{2}+\eta ^{\ast }k^{2}\right) w_{\mathbf{k}\perp }=0,
\label{5.6}
\end{equation}
and a coupled set of equations for $\rho _{\mathbf{k}},\theta
_{\mathbf{k} },\Pi _{\mathbf{k}}$, and the longitudinal velocity
component $w_{\mathbf{k} ||}$ (parallel to $\mathbf{k}$)
\begin{equation}
\frac{\partial \delta z_{\mathbf{k}\alpha }(\tau )}{\partial \tau
}=\left( M_{\alpha \beta }^{(0)}+ikM_{\alpha \beta
}^{(1)}+k^{2}M_{\alpha \beta }^{(2)}\right) \delta
z_{\mathbf{k}\beta }(\tau ),  \label{5.7}
\end{equation}
where now $\delta z_{\mathbf{k}\alpha }(\tau )$ denotes the four
variables $ \left( \rho _{\mathbf{k}},\theta _{\mathbf{k}},\Pi
_{\mathbf{k}},w_{\mathbf{k }||}\right) $. The matrices in this
equation are
\begin{equation}
M^{(0)}=\left(
\begin{array}{cccc}
0 & 0 & 0 & 0 \\
-x_{1}\left( \frac{\partial \zeta ^{\ast }}{\partial x_{1}}\right)
_{T,p} &
\frac{\zeta ^{\ast }}{2} & -\zeta ^{\ast } & 0 \\
-x_{1}\left( \frac{\partial \zeta ^{\ast }}{\partial x_{1}}\right)
_{T,p} &
\frac{\zeta ^{\ast }}{2} & -\zeta ^{\ast } & 0 \\
0 & 0 & 0 & \frac{\zeta ^{\ast }}{2}
\end{array}
\right) ,  \label{5.8}
\end{equation}
\begin{equation}
M^{(1)}=\left(
\begin{array}{cccc}
0 & 0 & 0 & 0 \\
0 & 0 & 0 & -\frac{2}{3} \\
0 & 0 & 0 & -\frac{5}{3} \\
0 & 0 & -\frac{1}{2}\frac{\mu _{12}}{x_{1}\mu +x_{2}} & 0
\end{array}
\right) ,  \label{5.9}
\end{equation}
\begin{equation}
M^{(2)}=\left(
\begin{array}{cccc}
-\frac{1}{2}\frac{\mu x_{1}+x_{2}}{1+\mu }D^{\ast } &
-\frac{1}{2x_{1}}\frac{ \mu x_{1}+x_{2}}{1+\mu }D^{\prime
}{}^{\ast } & -\frac{1}{2x_{1}}\frac{\mu
x_{1}+x_{2}}{1+\mu }D_{p}^{\ast } & 0 \\
-x_{1}\left( \frac{2}{3}D^{\prime \prime }{}^{\ast }-\frac{1-\mu
}{2(1+\mu )} D^{\ast }\right)  & \frac{1-\mu }{2(1+\mu )}D^{\prime
}{}^{\ast }-\frac{2}{3} \lambda ^{\ast } & -\frac{2}{3}L^{\ast
}+\frac{1-\mu }{2(1+\mu )}D_{p}^{\ast
} & 0 \\
-\frac{2}{3}x_{1}D^{\prime \prime }{}^{\ast} & -\frac{2}{3}
\lambda ^{\ast } & -\frac{2}{3}L^{\ast } & 0 \\
0 & 0 & 0 & -\frac{4}{3}\eta ^{\ast }
\end{array}
\right) .  \label{5..10}
\end{equation}
In these equations, $x_{i}=n_{iH}/n_{H}$, $\zeta ^{\ast }=\zeta
_{H}/\nu _{H} $ and the reduced transport coefficients $D^{\ast
}$, $ D_{p}^{\ast }$, and $D^{\prime }{}^{\ast }$ are given by
Eqs.\ (\ref{2.9})--(\ref{2.11} ), respectively. Moreover, the
reduced Navier--Stokes transport coefficients are
\begin{equation}
\eta ^{\ast }=\frac{\nu _{H}\eta }{\rho _{H}v_{0H}^{2}},
\label{5.11}
\end{equation}
\begin{equation}
D^{\prime \prime }{}^{\ast }=\frac{\nu _{H}T_{H}D^{\prime \prime
}}{ n_{H}v_{0H}^{2}},\quad L^{\ast }=\frac{\nu
_{H}L}{v_{0H}^{2}},\quad \lambda ^{\ast }=\frac{\nu _{H}\lambda
}{n_{H}v_{0H}^{2}},  \label{5.12}
\end{equation}
where $\rho_H=m_1n_{1H}+m_2n_{2H}$.

It is instructive to consider first the solutions to these
equations in the extreme long wavelength limit, $k=0$. In this
case, the eigenvalues or hydrodynamic modes are given by
\begin{equation}
s_{\perp }=\frac{1}{2}\zeta ^{\ast },\hspace{0.3in}s_{n}=\left(
0,0,-\frac{1}{2} \zeta ^{\ast },\frac{1}{2}\zeta ^{\ast }\right),
\label{5.14}
\end{equation}
where $s_n$ refers to the longitudinal modes. Two of the
eigenvalues are positive, corresponding to growth of the initial
perturbation in time. Thus, some of the solutions are unstable.
The two zero eigenvalues represent marginal stability solutions,
while the negative eigenvalue gives stable solutions. For general
initial perturbations all modes are excited. These modes
correspond to evolution of the fluid due to uniform perturbations
of the HCS, i.e. a global change in the HCS parameters. The
unstable modes are seen to arise from the initial perturbations
$w_{\mathbf{k}\perp }(0)$ or $w_{\mathbf{k}||}(0)$. The marginal
modes correspond to changes in the composition at fixed pressure,
density, and velocity, and to changes in $\Pi_{{\bf
k}}-\theta_{{\bf k}}$ at constant composition and velocity. The
decaying mode corresponds to changes in the temperature or
pressure for $\Pi_{{\bf k}}=\theta_{{\bf k}}$. The unstable modes
may appear trivial since they are due entirely to the
normalization of the fluid velocity by the time dependent thermal
velocity. However, this normalization is required by the scaling
of the entire set of equations to obtain time independent
coefficients.

\begin{figure}[tbp]
\begin{center}
\resizebox{7.5cm}{!}{\includegraphics{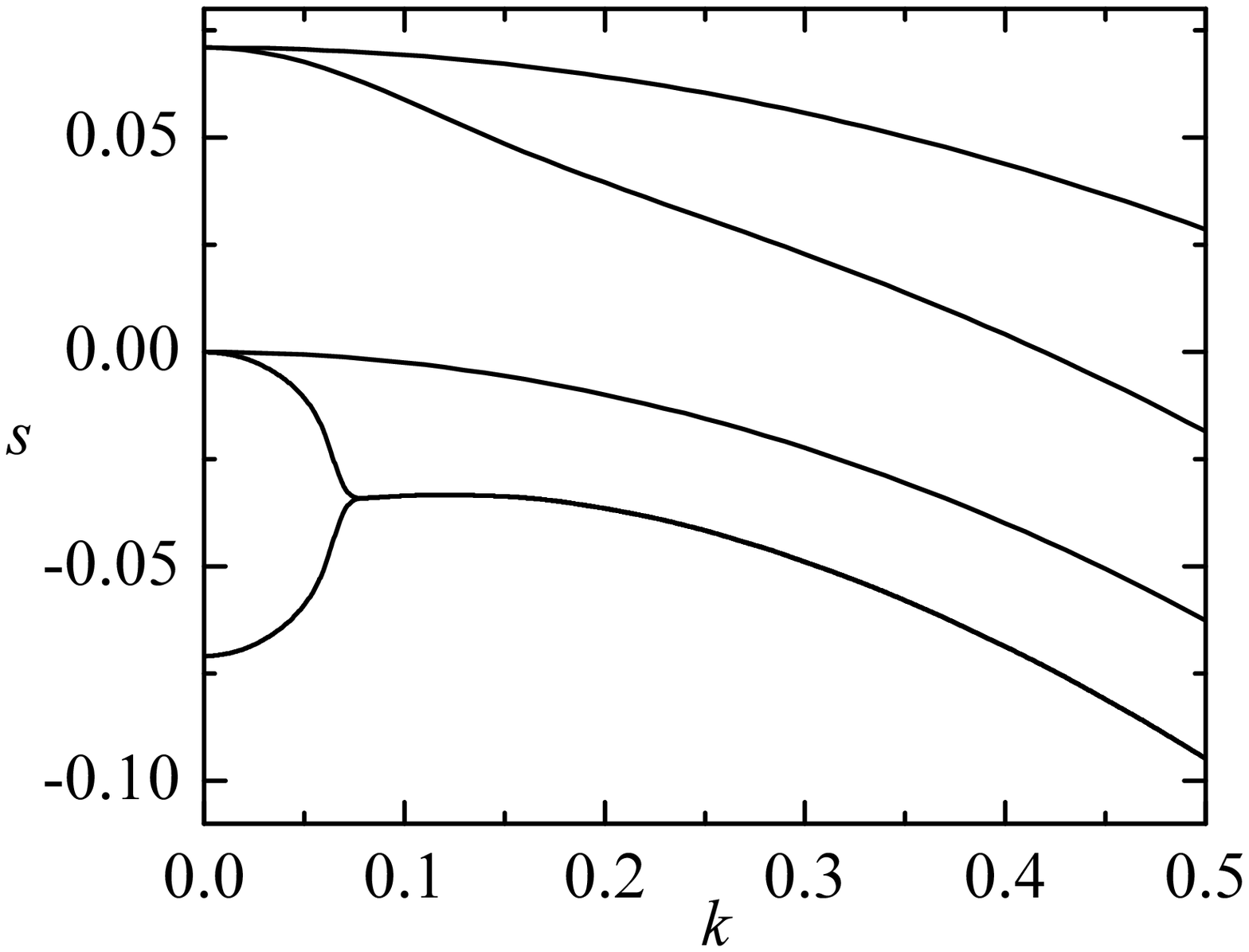}}
\end{center}
\caption{Dispersion relations for $\protect\alpha =0.9$,
$x_{1}=0.2$, $ \protect\omega =1$ and $\protect\mu =4$.}
\label{fig10}
\end{figure}

At finite wave vectors, these instabilities give rise to real
growth of spatial perturbations. The linear growth of the
transverse modes is simply given by
\begin{equation}
w_{\mathbf{k}\perp }(\tau )=w_{\mathbf{k}\perp }(0)\exp
(\frac{1}{2}\zeta ^{\ast }-\eta ^{\ast }k^{2})\tau .  \label{5.23}
\end{equation}
The instability for the two shear modes is removed at sufficiently
large $ k>k_{\perp }^{\text{c}}$, where
\begin{equation}
k_{\perp }^{\text{c}}=\left( \frac{\zeta ^{\ast }}{2\eta ^{\ast
}}\right) ^{1/2}.  \label{5.25}
\end{equation}
The wave vector dependence of the remaining four modes is more
complex. This is illustrated in Fig.\ \ref{fig10} showing the real
parts of the modes $ s\left( k\right) $ for $\alpha _{ij}=0.9,$
$\sigma _{1}/\sigma _{2}=1$, $ x_{1}=0.2$, and $m_{1}/m_{2}=4$.
The $k=0$ values are those of (\ref{5.14}), corresponding to six
hydrodynamic modes with two different degeneracies. The shear mode
degeneracy remains at finite $k$ but the other is removed at any
finite $k$. At sufficiently large $k$ a pair of real modes become
equal and become a complex conjugate pair at all larger wave
vectors, like a sound mode. The smaller of the unstable modes is
that associated with the longitudinal velocity, which couples to
the scalar hydrodynamic fields. It becomes negative at a wave
vector smaller than that of Eq.\ (\ref{5.25}) and gives the
threshold for development of spatial instabilities.

The results obtained here for the mixture show no new surprises
relative to the earlier work for a one component gas,
\cite{BDKS98,G05} with only the addition of the stable mass
diffusion mode. Of course, the quantitative features can be quite
different since there are additional degrees of freedom with the
parameter set $\left\{ x_{1},T,m_{1}/m_{2},\sigma _{1}/\sigma
_{2},\alpha _{ij}\right\} .$ Also, the manner in which these
linear instabilities are enhanced by the nonlinearities may be
different from that for the one component case.

\section{Summary and Discussion}
\label{sec6}

The Navier--Stokes order hydrodynamic equations have been
discussed for a low density granular binary mixture. The form of
the momentum flux is the same as for a one component gas, with
only the value of the viscosity changed (see Appendix \ref{appA}).
Since the dependence of the viscosity on the parameters of the
mixture has been widely explored in a previous paper, \cite{MG03}
attention has been focused here on the mass and heat fluxes and
their associated transport coefficients. There is no phenomenology
involved as the equations and the transport coefficients have been
derived systematically from the inelastic Boltzmann equation by
the Chapman-Enskog procedure. Consequently, there is no \emph{a
priori} limitation on the degree of inelasticity, size and mass
ratios, or composition. For practical purposes, the integral
equations determining the transport coefficients have been solved
by truncated expansions in Sonine polynomials. This is expected to
fail at extreme values of size or mass ratio, \cite{MG03,GM04} but
the results are quite accurate otherwise.

The hydrodynamic equations are the same as for a normal gas,
except for a sink in the energy equation due to granular cooling,
and additional transport coefficients in the mass and heat flux
constitutive equations. The latter arise because the usual
restrictions of irreversible thermodynamics no longer apply. These
restrictions include Onsager reciprocal relations among various
transport coefficients, and the extent to which these are violated
has been demonstrated in Section 4. It has been verified that the
results described here reduce to those for a normal mixture in the
elastic limit, \cite{CC70} and to those for a one component
granular gas \cite{BDKS98} when the species are mechanically
identical.

As is the case for a normal gas, the hydrodynamic fields include
only the global temperature even though two species temperatures
can be defined. For a normal gas the species temperatures rapidly
approach the global temperature due to equipartition. For the
granular gas the species temperatures approach different values,
but with the same time dependence as the global temperature. The
transport coefficients have an additional dependence on the
composition due to this time independent ratio of species
temperatures. This has been illustrated in Fig.\ \ref{fig3} for
the thermal diffusion coefficient, where the effect is seen to be
large for large mass ratio (implying very different species
temperatures. Generally, it was seen that the deviation due to
inelasticity is enhanced for greater mechanical differences
between the species.

The linear equations for small perturbations of the special
homogeneous cooling solution to the hydrodynamic equations were
obtained and discussed. In order to characterize the solutions in
terms of modes, it is necessary to introduce dimensionless fields
so that the time dependence of the reference HCS is eliminated.
The resulting equations exhibit a long wavelength instability for
three of the modes. This is quite similar to the case of a one
component granular gas, \cite{BDKS98,G05} and in fact the same
modes are unstable here. The additional diffusion mode for two
species behaves as for a normal fluid. The consequences of this
instability for a binary mixture were not studied here. This
entails an analysis of the dominant nonlinearities which has not
been performed as yet. Since there are additional degrees of
freedom would be interesting to see if the density clustering that
occurs for a one component system is more complex here (e.g.,
species segregation).

\acknowledgments

Partial support of the Ministerio de Ciencia y Tecnolog\'{\i}a
(Spain) through Grant No.  FIS2004-01399 (partially financed by
FEDER funds) in the case of V.G. and ESP2003-02859 (partially
financed by FEDER funds) in the case of J.M.M. is acknowledged. V.
G. also acknowledges support from the European Community's Human
Potential Programme HPRN-CT-2002-00307 (DYGLAGEMEM).

\appendix
\section{Some explicit expressions}
\label{appA}

Navier--Stokes hydrodynamics retains terms up through second order
in the gradients. As a scalar, the cooling rate has the most
general form at this order given by
\begin{eqnarray}
\zeta  &=&\zeta _{0}+\zeta _{u}\nabla \cdot \mathbf{u}+\zeta
_{x}\nabla ^{2}x_{1}+\zeta _{T}\nabla ^{2}T+\zeta _{p}\nabla
^{2}p+\zeta _{TT}\left( \nabla T\right) ^{2}+\zeta _{xx}\left(
\nabla x_{1}\right) ^{2}\nonumber\\
& &  +\zeta _{pp}\left( \nabla p\right) ^{2}+\zeta _{Tx}\left(
\mathbf{\nabla }T\right) \cdot \left( \mathbf{\nabla }x_{1}\right)
+\zeta _{Tp}\left( \mathbf{\nabla }T\right) \cdot \left(
\mathbf{\nabla } p\right) +\zeta _{px}\left( \mathbf{\nabla
}p\right) \cdot \left( \mathbf{ \nabla }x_{1}\right) \nonumber\\
& & +\zeta _{uu}\left( \nabla _{i}u_{j}\right) \left( \nabla
_{i}u_{j}\right) .  \label{a01}
\end{eqnarray}
For a low density gas the first order gradient term vanishes,
\cite{BDKS98} $\zeta _{u}=0$. However, for higher densities $\zeta
_{u}$ is different from zero. \cite{GD99a} The second order terms
have been left implicit in equations (\ref{2.5}) and (\ref{2.6})
for the temperature and the pressure. As noted in the text, these
second order terms have been calculated for a one component fluid
\cite{BDKS98} and found to be very small relative to corresponding
terms from the fluxes. Consequently, they have been neglected in
the linearized hydrodynamic equations (\ref{5.7}).

\subsection{Mass flux parameters}

The transport coefficients are expressed in terms of a number of
dimensionless parameters. For completeness, they are listed here.
In general they depend on the reference distribution functions in
the Chapman--Enskog expansion, which are not Maxwellians.
\cite{GD99} The deviation of the reference distributions from
their Maxwellian forms is measured by the cumulants $c_i$.
\cite{GD99} As shown in Sec.\ \ref{sec3}, while the influence of
these coefficients on the transport coefficients is not quite
important in the case of the coefficients associated with the mass
flux (see Fig.\ \ref{fig1}, for example), not happens the same in
the case of the heat flux (see Fig.\ \ref{fig4}, for example)
where the influence of $c_i$ is not negligible for strong
dissipation. However, in order to offer a simplified theory the
parameters given in this Appendix have neglected the corrections
due to the cumulants $c_i$. The full expressions for the transport
coefficients can be found in Ref.\ \onlinecite{GD02}.

The temperature ratio $\gamma =T_{1}/T_{2}$ is determined from the
condition
\begin{equation}
\label{n1} \zeta_1^*=\zeta_2^*=\zeta^*,
\end{equation}
where the dimensionless cooling rate, $\zeta_i ^{\ast}=\zeta_i
/\nu _{0}$ ($\nu _{0}$ is the average frequency defined below
(\ref{2.8})) is
\begin{eqnarray}
\zeta_1^{\ast } &=&\frac{2}{3}\sqrt{2\pi }\left( \frac{\sigma
_{1}}{\sigma _{12}}\right) ^{2}x_{1}\theta _{1}^{-1/2}\left(
1-\alpha _{11}^{2}\right)
\nonumber \\
&&+\frac{4}{3}\sqrt{\pi }x_{2}\mu _{21}\left( \frac{1+\theta
}{\theta }\right) ^{1/2}\left( 1+\alpha _{12}\right) \theta
_{2}^{-1/2}\left[ 2-\mu _{21}\left( 1+\alpha _{12}\right) \left(
1+\theta \right) \right]. \label{a03}
\end{eqnarray}
The expression for $\zeta_2^*$ can be easily obtained by
interchanging $1\leftrightarrow 2$. Here, $\theta _{1}=1/(\mu
_{21}\gamma _{1})$, $\theta _{2}=1/(\mu _{12}\gamma _{2})$, $\mu
_{ij}=m_{i}/(m_{i}+m_{j})$, $\delta =x_{1}/x_{2}$, and $\theta
=\theta _{1}/\theta _{2}=\mu /\gamma $. The temperature ratios
$\gamma_i$ are related to the temperature ratio $\gamma$ through
Eqs.\ (\ref{2.11a}). In the quasielastic limit ($\alpha_{ij}\ll
1$), the temperature ratio $\gamma$ has the simple form
\begin{eqnarray}
\gamma  &\rightarrow&1+\frac{1}{2\mu _{12}\mu _{21}}\left\{ \left(
\mu _{12}x_{1}-\mu
_{21}x_{2}\right) \left( 1-\alpha _{12}\right) \right.   \nonumber \\
&&\left. +\frac{1}{\sqrt{2}}\left[ \left( \frac{\sigma
_{22}}{\sigma _{12}} \right) ^{2}x_{2}\sqrt{\mu _{12}}\left(
1-\alpha _{22}\right) -\left( \frac{ \sigma _{11}}{\sigma
_{12}}\right) ^{2}x_{1}\sqrt{\mu _{21}}\left( 1-\alpha
_{11}\right) \right] \right\} .  \label{a02}
\end{eqnarray}
In this limit, the temperature ratio is a linear function of the
hydrodynamic field $x_{1}$. The dimensionless frequency $\nu
^{\ast}$ appearing in the expressions of the transport
coefficients associated with the mass flux is
\begin{equation}
\nu ^{\ast }=\frac{4}{3}\mu _{21}\frac{1+\mu \delta }{1+\delta
}\left( \frac{ 1+\theta }{\theta }\right) ^{1/2}\theta
_{2}^{-1/2}(1+\alpha _{12}). \label{a04}
\end{equation}
The cooling rate $\zeta ^{\ast}$ and frequency $\nu ^{\ast }$ are
also functions of the hydrodynamic field $x_{1}$. However, all the
parameters above are independent of the temperature and density.

\subsection{Heat and momentum flux parameters}

As in the last section effects due to the distortion of the
reference Maxwellian are neglected. In the case of the heat flux
$\mathbf{q}$, Eq.\ ( \ref{2.2}), the transport coefficients
$D^{\prime \prime }$, $L$, and $ \lambda $ are given by Eqs.\
(\ref{2.12})--(\ref{2.14}), respectively. By using matrix
notation, the (dimensionless) Sonine coefficients $ d_{i}^{\prime
\prime }$, $\ell _{i}$, and $\lambda _{i}$ verify the coupled set
of six equations \cite{GD02}
\begin{equation}
\Lambda _{\sigma \sigma ^{\prime }}X_{\sigma ^{\prime }}=Y_{\sigma
}. \label{a1}
\end{equation}
where $X_{\sigma ^{\prime }}$ is the column matrix
\begin{equation}
\mathbf{X}=\left(
\begin{array}{c}
d_{1}^{\prime \prime } \\
d_{2}^{\prime \prime } \\
\ell _{1} \\
\ell _{2} \\
\lambda _{1} \\
\lambda _{2}
\end{array}
\right) ,  \label{a2}
\end{equation}
and $\Lambda _{\sigma \sigma ^{\prime }}$ is the matrix
\begin{equation}
\Lambda =\left(
\begin{array}{cccccc}
\nu _{11}-\frac{3}{2}\zeta ^{\ast } & \nu _{12} & -\left(
\frac{\partial \zeta ^{\ast }}{\partial x_{1}}\right) _{p,T} & 0 &
-\left( \frac{\partial
\zeta ^{\ast }}{\partial x_{1}}\right) _{p,T} & 0 \\
\nu _{21} & \nu _{22}-\frac{3}{2}\zeta ^{\ast } & 0 & -\left(
\frac{\partial \zeta ^{\ast }}{\partial x_{1}}\right) _{p,T} & 0 &
-\left( \frac{\partial
\zeta ^{\ast }}{\partial x_{1}}\right) _{p,T} \\
0 & 0 & \nu _{11}-\frac{5}{2}\zeta ^{\ast } & \nu _{12} & -\zeta
^{\ast } & 0
\\
0 & 0 & \nu _{21} & \nu _{22}-\frac{5}{2}\zeta ^{\ast } & 0 &
-\zeta ^{\ast }
\\
0 & 0 & \zeta ^{\ast }/2 & 0 & \nu _{11}-\zeta ^{\ast } & \nu _{12} \\
0 & 0 & 0 & \zeta ^{\ast }/2 & \nu _{21} & \nu _{22}-\zeta ^{\ast
}
\end{array}
\right) .  \label{a3}
\end{equation}
The column matrix $Y_{\sigma }$ has the elements
\begin{equation}
Y_{1}=D^{\ast }\left( \tau _{12}-\frac{\zeta ^{\ast }}{x_{1}\gamma
_{1}^{2}} \right) -\frac{1}{\gamma _{1}^{2}}\left( \frac{\partial
\gamma _{1}}{
\partial x_{1}}\right) _{p,T},\hspace{0.3in}Y_{2}=-D^{\ast }\left( \tau
_{21}-\frac{\zeta ^{\ast }}{x_{2}\gamma _{2}^{2}}\right)
-\frac{1}{\gamma _{2}^{2}}\left( \frac{\partial \gamma
_{2}}{\partial x_{1}}\right) _{p,T}, \label{a5}
\end{equation}
\begin{equation}
Y_{3}=D_{p}^{\ast }\left( \tau _{12}-\frac{\zeta ^{\ast
}}{x_{1}\gamma _{1}^{2}}\right) ,\hspace{0.3in}Y_{4}=-D_{p}^{\ast
}\left( \tau _{21}-\frac{ \zeta ^{\ast }}{x_{2}\gamma
_{2}^{2}}\right) ,  \label{a7}
\end{equation}
\begin{equation}
Y_{5}=-\frac{1}{\gamma _{1}}+D^{\prime }{}^{\ast }\left( \tau
_{12}-\frac{ \zeta ^{\ast }}{x_{1}\gamma _{1}^{2}}\right)
,\hspace{0.3in}Y_{6}=-\frac{1}{ \gamma _{2}}-D^{\prime }{}^{\ast
}\left( \tau _{21}-\frac{\zeta ^{\ast }}{ x_{2}\gamma
_{2}^{2}}\right) .  \label{a9}
\end{equation}
The dimensionless collision integrals $\tau _{12}$, $\nu _{11}$,
and $\nu _{12}$ are given, respectively, by
\begin{eqnarray}
\tau _{12} &=&\frac{4}{3}\sqrt{\frac{\mu _{21}}{2}}\left(
\frac{\sigma _{1}}{ \sigma _{12}}\right) ^{2}\gamma
_{1}^{-3/2}(1-\alpha _{11}^{2}) \nonumber
\label{a11} \\
&&+\frac{4}{15}\mu _{21}^{-1}\gamma _{1}^{-4}(1+\alpha
_{12})\left( \theta _{1}+\theta _{2}\right) ^{-1/2}\left( \theta
_{1}\theta _{2}\right) ^{-3/2}\left( \frac{x_{2}}{x_{1}}A-\gamma
B\right) ,
\end{eqnarray}
\begin{eqnarray}
\label{a12}
 \nu _{11} &=&\frac{16}{15}x_{1}\left( \frac{\sigma
_{1}}{\sigma _{12}} \right) ^{2}(2\theta _{1})^{-1/2}(1+\alpha
_{11})\left[ 1+\frac{33}{16}
(1-\alpha _{11})\right]   \nonumber \\
&&+\frac{2}{15}x_{2}\mu _{21}(1+\alpha _{12})\left( \frac{\theta
_{1}}{ \theta _{2}(\theta _{1}+\theta _{2})}\right) ^{3/2}\left(
E-5\frac{\theta _{1}+\theta _{2}}{\theta _{1}}A\right) ,
\end{eqnarray}
\begin{equation}
\label{a13}
 \nu _{12}=-\frac{2}{15}x_{2}\frac{\mu _{21}^{2}}{\mu
_{12}}(1+\alpha _{12})\left( \frac{\theta _{1}}{\theta _{2}(\theta
_{1}+\theta _{2})}\right) ^{3/2}\left( F+5\frac{\theta _{1}+\theta
_{2}}{\theta _{2}}B\right) .
\end{equation}
In the above equations we have introduced the quantities
\cite{misprints}
\begin{eqnarray}
\label{a14} A &=&5(2\beta _{12}+\theta _{2})+\mu _{21}(\theta
_{1}+\theta _{2})\left[ 5(1-\alpha _{12})-2(7\alpha _{12}-11)\beta
_{12}\theta _{1}^{-1}\right]
\nonumber \\
&&+18\beta _{12}^{2}\theta _{1}^{-1}+2\mu _{21}^{2}\left( 2\alpha
_{12}^{2}-3\alpha _{12}+4\right) \theta _{1}^{-1}(\theta
_{1}+\theta _{2})^{2}-5\theta _{2}\theta _{1}^{-1}(\theta
_{1}+\theta _{2})
\end{eqnarray}
\begin{eqnarray}
\label{a15} B &=&5(2\beta _{12}-\theta _{1})+\mu _{21}(\theta
_{1}+\theta _{2})\left[ 5(1-\alpha _{12})+2(7\alpha _{12}-11)\beta
_{12}\theta _{2}^{-1}\right]
\nonumber \\
&&-18\beta _{12}^{2}\theta _{2}^{-1}-2\mu _{21}^{2}\left( 2\alpha
_{12}^{2}-3\alpha _{12}+4\right) \theta _{2}^{-1}(\theta
_{1}+\theta _{2})^{2}+5(\theta _{1}+\theta _{2})
\end{eqnarray}
\begin{eqnarray}
\label{a16} E &=&2\mu _{21}^{2}\theta _{1}^{-2}(\theta _{1}+\theta
_{2})^{2}\left( 2\alpha _{12}^{2}-3\alpha _{12}+4\right) (5\theta
_{1}+8\theta _{2})
\nonumber \\
&&-\mu _{21}(\theta _{1}+\theta _{2})\left[ 2\beta _{12}\theta
_{1}^{-2}(5\theta _{1}+8\theta _{2})(7\alpha _{12}-11)+2\theta
_{2}\theta
_{1}^{-1}(29\alpha _{12}-37)-25(1-\alpha _{12})\right]   \nonumber \\
&&+18\beta _{12}^{2}\theta _{1}^{-2}(5\theta _{1}+8\theta
_{2})+2\beta
_{12}\theta _{1}^{-1}(25\theta _{1}+66\theta _{2})  \nonumber \\
&&+5\theta _{2}\theta _{1}^{-1}(11\theta _{1}+6\theta
_{2})-5(\theta _{1}+\theta _{2})\theta _{1}^{-2}\theta
_{2}(5\theta _{1}+6\theta _{2})
\end{eqnarray}
\begin{eqnarray}
\label{a17} F &=&2\mu _{21}^{2}\theta _{2}^{-2}(\theta _{1}+\theta
_{2})^{2}\left( 2\alpha _{12}^{2}-3\alpha _{12}+4\right) (8\theta
_{1}+5\theta _{2})
\nonumber \\
&&-\mu _{21}(\theta _{1}+\theta _{2})\left[ 2\beta _{12}\theta
_{2}^{-2}(8\theta _{1}+5\theta _{2})(7\alpha _{12}-11)-2\theta
_{1}\theta
_{2}^{-1}(29\alpha _{12}-37)+25(1-\alpha _{12})\right]   \nonumber \\
&&+18\beta _{12}^{2}\theta _{2}^{-2}(8\theta _{1}+5\theta
_{2})-2\beta
_{12}\theta _{2}^{-1}(66\theta _{1}+25\theta _{2})  \nonumber \\
&&+5\theta _{1}\theta _{2}^{-1}(6\theta _{1}+11\theta
_{2})-5(\theta _{1}+\theta _{2})\theta _{2}^{-1}(6\theta
_{1}+5\theta _{2})
\end{eqnarray}
Here, $\beta _{12}=\mu _{12}\theta _{2}-\mu _{21}\theta _{1}$. The
corresponding expressions for $\tau _{21}$, $\nu _{22}$, and $\nu
_{21}$ can be inferred from Eqs.\ (\ref{a11})--(\ref{a17}) by
interchanging $ 1\leftrightarrow 2$. For elastic collisions, the
expressions (\ref{a11})--( \ref{a17}) reduce to those obtained for
hard sphere mixtures. \cite{CC70bis}

The solution to Eq.\ (\ref{a1}) is
\begin{equation}
X_{\sigma }=\left( \Lambda ^{-1}\right) _{\sigma \sigma ^{\prime
}}Y_{\sigma ^{\prime }}.  \label{a18.0}
\end{equation}
This relation provides an explicit expression for the coefficients
$ d_{i}^{\prime \prime }$, $\ell _{i}$, and $\lambda _{i}$ in
terms of the coefficients of restitution  and the parameters of
the mixture. Their explicit forms are
\begin{eqnarray}
\label{a18} d_1^{\prime \prime}&=&\frac{1}{\Delta}\left\{2\left[2
\nu_{12}Y_2-Y_1(2\nu_{22}-3\zeta^*)\right]\left[\nu_{12}\nu_{21}-\nu_{11}\nu_{22}
+2(\nu_{11}+\nu_{22})\zeta^*-4\zeta^{*2}\right]\right.\nonumber\\
&& +2\left( \frac{\partial \zeta ^{\ast }}{\partial x_{1}}\right)
_{p,T}(Y_3+Y_5)\left[2\nu_{12}\nu_{21}+2\nu_{22}^2-\zeta^*(
7\nu_{22}-6\zeta^{*})\right]\nonumber\\
& & \left.-2\nu_{12}\left( \frac{\partial \zeta ^{\ast }}{\partial
x_{1}}\right)_{p,T}(Y_4+Y_6)\left(2\nu_{11}+2\nu_{22}-7\zeta^*\right)\right\},
\end{eqnarray}
\begin{eqnarray}
\label{a19} \ell_1&=&\frac{1}{\Delta}\left\{-2Y_3\left[2
(\nu_{12}\nu_{21}-\nu_{11}\nu_{22})\nu_{22}+\zeta^*(7\nu_{11}\nu_{22}-5\nu_{12}\nu_{21}+2\nu_{22}^2
-6\nu_{11}\zeta^*-7\nu_{22}\zeta^*+6\zeta^{*2})\right]\right.\nonumber\\
&&
+2Y_4\nu_{12}\left[2\nu_{12}\nu_{21}-2\nu_{11}\nu_{22}+2\zeta^*(\nu_{11}+\nu_{22})
-\zeta^{*2}\right]\nonumber\\
& &
\left.+2Y_5\zeta^*\left[2\nu_{12}\nu_{21}+\nu_{22}(2\nu_{22}-7\zeta^*)+6\zeta^{*2}\right]
-2\nu_{12}\zeta^*Y_6\left[2(\nu_{11}+\nu_{22})-7\zeta^*\right]
\right\},
\end{eqnarray}
\begin{eqnarray}
\label{a20} \lambda_1&=&\frac{1}{\Delta}\left\{-Y_3\zeta^*\left[2
\nu_{12}\nu_{21}+\nu_{22}(2\nu_{22}-7\zeta^*)+6\zeta^{*2}\right]+\nu_{12}\zeta^*
Y_4\left[2(\nu_{11}+\nu_{22})-7\zeta^*\right] \right. \nonumber\\
&
&-Y_5\left[4\nu_{12}\nu_{21}(\nu_{22}-\zeta^*)+2\nu_{22}^2(5\zeta^*-2\nu_{11})+2\nu_{11}
(7\nu_{22}\zeta^*-6\zeta^{*2})+5\zeta^{*2}(6\zeta^*-7\nu_{22})\right]\nonumber\\
& & \left.
+\nu_{12}Y_6\left[4\nu_{12}\nu_{21}+2\nu_{11}(5\zeta^*-2\nu_{22})+\zeta^*(10\nu_{22}-
23\zeta^*)\right]\right\},
\end{eqnarray}
where
\begin{equation}
\label{a21}
\Delta=\left[4(\nu_{12}\nu_{21}-\nu_{11}\nu_{22})+6\zeta^*(\nu_{11}+\nu_{22})-9\zeta^{*2}\right]
\left[\nu_{12}\nu_{21}-\nu_{11}\nu_{22}+2\zeta^*(\nu_{11}+\nu_{22})-4\zeta^{*2}\right].
\end{equation}
The expressions for $d_2^{\prime \prime}$, $\ell_2$, and
$\lambda_2$ can be obtained from Eqs.\ (\ref{a18})--(\ref{a20}) by
setting $1\leftrightarrow 2$. From the above expressions one can
easily get the transport coefficients $D^{\prime }$, $L$ and
$\lambda $ from Eqs.\ (\ref{2.12})--(\ref{2.14}), respectively.
They are functions of $x_{1}$ but independent of temperature and
pressure.

The pressure tensor $P_{k,\ell }$ is given by
\begin{equation}
P_{k\ell }=p\delta _{k\ell }-\eta \left( \nabla _{\ell
}u_{k}+\nabla _{k}u_{\ell }-\frac{2}{3}\delta _{k\ell }\nabla
\cdot \mathbf{u}\right) , \label{a22}
\end{equation}
where $\eta $ is the shear viscosity coefficient. Its expression
can be written as \cite{GD02}
\begin{equation}
\eta =\frac{nT}{\nu _{0}}\left( x_{1}\gamma _{1}^{2}\eta
_{1}+x_{2}\gamma _{2}^{2}\eta _{2}\right) ,  \label{a22.1}
\end{equation}
with
\begin{equation}
\eta _{1}=\frac{2\gamma _{2}(2\lambda _{22}-\zeta ^{\ast
})-4\gamma _{1}\lambda _{12}}{\gamma _{1}\gamma _{2}\left[ \zeta
^{\ast 2}-2\zeta ^{\ast }(\lambda _{11}+\lambda _{22})+4(\lambda
_{11}\lambda _{22}-\lambda _{12}\lambda _{21})\right] },
\label{a23}
\end{equation}
\begin{equation}
\eta _{2}=\frac{2\gamma _{1}(2\lambda _{11}-\zeta ^{\ast
})-4\gamma _{2}\lambda _{21}}{\gamma _{1}\gamma _{2}\left[ \zeta
^{\ast 2}-2\zeta ^{\ast}(\lambda _{11}+\lambda _{22})+4(\lambda
_{11}\lambda _{22}-\lambda _{12}\lambda _{21})\right] }.
\label{a24}
\end{equation}
The dimensionless quantities $\lambda _{ij}$ are given by
\cite{GD02,MG03}
\begin{eqnarray}
\lambda _{11} &=&\frac{16}{5\sqrt{2}}x_{1}\left( \frac{\sigma
_{1}}{\sigma _{12}}\right) ^{2}\theta _{1}^{-1/2}\left[
1-\frac{1}{4}(1-\alpha _{11})^{2}
\right]   \nonumber  \label{a25} \\
&&+\frac{8}{15}x_{2}\mu _{21}(1+\alpha _{12})\theta
_{1}^{3/2}\theta _{2}^{-1/2}\left[ 6\theta _{1}^{-2}(\mu
_{12}\theta _{2}-\mu _{21}\theta
_{1})(\theta _{1}+\theta _{2})^{-1/2}\right.   \nonumber \\
&&\left.+\frac{3}{2}\mu _{21}\theta _{1}^{-2}(\theta _{1}+\theta
_{2})^{1/2}(3-\alpha _{12})+5\theta _{1}^{-1}(\theta _{1}+\theta
_{2})^{-1/2}\right],
\end{eqnarray}
\begin{eqnarray}
\lambda _{12} &=&\frac{8}{15}x_{2}\frac{\mu _{21}^{2}}{\mu
_{12}}(1+\alpha _{12})\theta _{1}^{3/2}\theta _{2}^{-1/2}\left[
6\theta _{2}^{-2}(\mu _{12}\theta _{2}-\mu _{21}\theta
_{1})(\theta _{1}+\theta
_{2})^{-1/2}\right.   \nonumber  \label{a26} \\
&&\left.+\frac{3}{2}\mu _{21}\theta _{2}^{-2}(\theta _{1}+\theta
_{2})^{1/2}(3-\alpha _{12})-5\theta _{2}^{-1}(\theta _{1}+\theta
_{2})^{-1/2}\right].
\end{eqnarray}
The corresponding expressions for $\lambda _{22}$ and $\lambda
_{21}$ can be inferred from Eqs.\ (\ref{a23}) and (\ref{a24}) by
interchanging $ 1\leftrightarrow 2$.

The program for calculating the cooling rates, the temperature
ratio and the transport coefficients of the binary mixture can be
obtained on request from the authors.


\begin{thebibliography}{99}
\bibitem{comment} Note that this coefficient is not strictly zero in
the approximation used here due to the fact that while the set
$\{d_{i}'', \ell_{i}, \lambda_{i}\}$ has been evaluated in the
second Sonine approximation, the coefficients $D^*$, $D_p^*$, and
$D'^*$ have been estimated in the first Sonine approximation.
However, the magnitude of $L_p$ is negligible.
\end{thebibliography}
\end{document}